\documentclass[10pt,journal,compsoc]{IEEEtran}
\ifCLASSOPTIONcompsoc
  \usepackage[nocompress]{cite}
\else
  \usepackage{cite}
\fi

\usepackage{cite}
\usepackage{amsmath,amssymb,amsfonts}
\usepackage{bbm}
\usepackage{graphicx}
\usepackage{mdframed}
\usepackage{textcomp}
\usepackage{xcolor}
\usepackage{kotex}
\usepackage{algorithmicx,algpseudocode}
\usepackage{setspace}
\usepackage{slashbox}
\usepackage{amsthm,thmtools,thm-restate}
\usepackage{subfigure}
\usepackage{booktabs}
\usepackage[linesnumbered,ruled]{algorithm2e}
\usepackage{multicol}
\usepackage{multirow}
\usepackage{verbatim} 
\usepackage[normalem]{ulem}
\usepackage{lettrine}

\usepackage{amsmath,amsfonts,amssymb}
\usepackage{tabularx}
\usepackage{tikz}

\declaretheorem[name=Lemma]{lemma}

\newcommand{\BfPara}[1]{{\noindent\bf#1.}\xspace}

\newcommand{\bra}[1]{\ensuremath{\left\langle#1\right|}}
\newcommand{\ket}[1]{\ensuremath{\left|#1\right\rangle}}
\newcommand{\thetab}[0]{\ensuremath{\boldsymbol{\theta}}}

\DeclareMathOperator*{\Motimes}{\raisebox{-0.25ex}{\scalebox{1.2}{$\bigotimes$}}}

\begin{document}
\title{Quantum Multi-Agent Reinforcement Learning for Cooperative Mobile Access in Space-Air-Ground Integrated Networks}

\author{
    Gyu Seon Kim, 
    Yeryeong Cho, 
    Jaehyun Chung, 
    Soohyun Park,~\IEEEmembership{Member,~IEEE}, 
    Soyi Jung,~\IEEEmembership{Member,~IEEE}, 
    Zhu Han,~\IEEEmembership{Fellow,~IEEE}, and
    Joongheon Kim,~\IEEEmembership{Senior Member,~IEEE}
        \IEEEcompsocitemizethanks{
        \IEEEcompsocthanksitem 
        G.S. Kim, Y. Cho, J. Chung, and J. Kim are with the School of Electrical Engineering, Korea University, Seoul 02841, Korea (e-mails: \{kingdom0545,joyena0909,rupang1234,joongheon\}@korea.ac.kr).
        \IEEEcompsocthanksitem 
        S. Park is with the Division of Computer Science, Sookmyung Women's University, Seoul 04310, Korea (e-mail: soohyun.park@sookmyung.ac.kr).
        \IEEEcompsocthanksitem 
        S. Jung is with the Department of Electrical and Computer Engineering, Ajou University, Suwon 16499, Korea (e-mail: sjung@ajou.ac.kr).
        \IEEEcompsocthanksitem 
        Z. Han is with the Department of Electrical and Computer Engineering, University of Houston, Texas, USA (e-mail: zhan2@uh.edu).
}
}

\IEEEtitleabstractindextext{
\begin{abstract}
Achieving global space-air-ground integrated network (SAGIN) access only with CubeSats presents significant challenges such as the access sustainability limitations in specific regions (e.g., polar regions) and the energy efficiency limitations in CubeSats. To tackle these problems, high-altitude long-endurance unmanned aerial vehicles (HALE-UAVs) can complement these CubeSat shortcomings for providing cooperatively global access sustainability and energy efficiency. However, as the number of CubeSats and HALE-UAVs, increases, the scheduling dimension of each ground station (GS) increases. As a result, each GS can fall into the curse of dimensionality, and this challenge becomes one major hurdle for efficient global access. Therefore, this paper provides a quantum multi-agent reinforcement Learning (QMARL)-based method for scheduling between GSs and CubeSats/HALE-UAVs in order to improve global access availability and energy efficiency. The main reason why the QMARL-based scheduler can be beneficial is that the algorithm facilitates a logarithmic-scale reduction in scheduling action dimensions, which is one critical feature as the number of CubeSats and HALE-UAVs expands. Additionally, individual GSs have different traffic demands depending on their locations and characteristics, thus it is essential to provide differentiated access services. The superiority of the proposed scheduler is validated through data-intensive experiments in realistic CubeSat/HALE-UAV settings.
\end{abstract}

}

\maketitle

\IEEEdisplaynontitleabstractindextext

\IEEEpeerreviewmaketitle

\begin{figure*}[t!]
  \centering
  \includegraphics[width=0.7\linewidth]{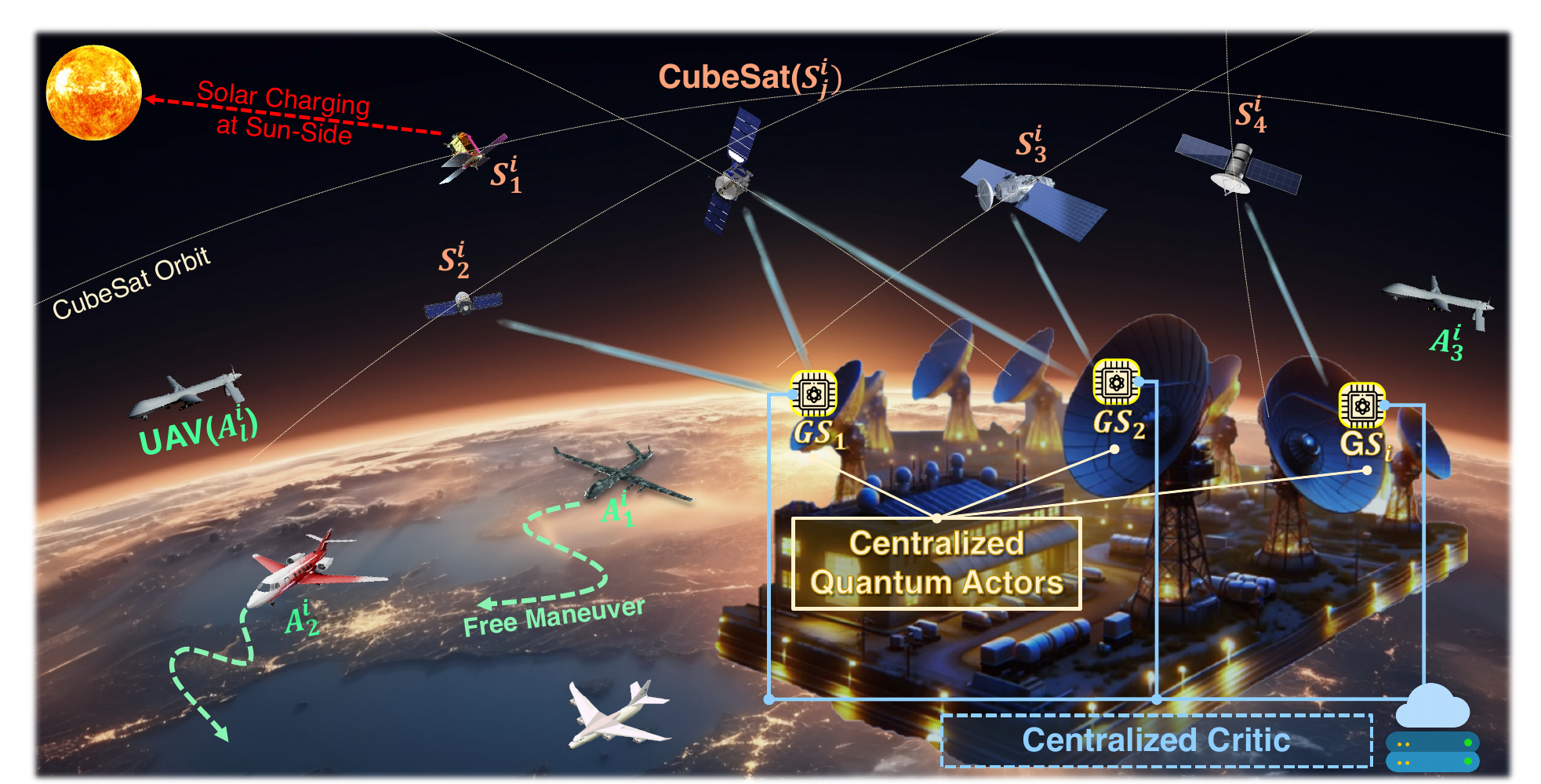}
  \caption{Reference network model.}
  \label{fig:Overview of this paper}
\end{figure*}

\section{Introduction}\label{sec:1}
Ultra-small-scale and low-cost cube satellites (CubeSats) have recently emerged as novel electrical aerospace devices in non-terrestrial networks (NTN) as one major component of global space-air-ground integrated network (SAGIN) systems in order to realize seamless global access services~\cite{TMC_satellite_4}. 
In the past, geostationary (GEO) satellites at the altitude of approximately $36,000$ km were employed for the global access services, yet their considerable distances from the Earth introduced extremely long propagation delays, which hindered the global access services~\cite{jsac_geo_pre}. Given that CubeSats operate as low Earth orbit (LEO) satellites at the altitude of approximately $500$\,km, they are more adept at facilitating global access services, offering reduced delays compared to GEO-based services~\cite{aimlab_LEO, TMC_satellite_2}.
However, the lower altitude of CubeSats, results in considerably smaller coverage compared to GEO-based services. Consequently, in order to achieve seamless global access, a significantly larger fleet of CubeSats is essentially required~\cite{TMC_satellite_6_CubeSat}. 
To take care of large-scale CubeSats, it is essentially required to design efficient scheduling algorithms for global access availability and energy efficiency. 
For more details, employing CubeSats to deliver global SAGIN mobile access necessitates determinations regarding \textit{which CubeSats should engage in the global access} amidst a scenario where a multitude of CubeSats are present.
This scenario culminates in a \textit{scheduling problem}, which can be conceptualized within the framework of multi-agent reinforcement learning (MARL)~\cite{jsac_marl}. The essence of this approach stems from the necessity for multiple ground stations (GSs) to collaboratively orchestrate the scheduling and servicing of their CubeSats to facilitate global SAGIN mobile access, as depicted in Fig.~\ref{fig:Overview of this paper}.
In the environment where multiple CubeSats exist, each GS cooperatively schedules CubeSats to participate in global SAGIN mobile access, and the corresponding efficient scheduling algorithms are needed.
Due to CubeSat's limited resources such as limited energy and bandwidth, without an efficient scheduling algorithm, it is impossible to optimally utilize these resources, maintain high quality of service (QoS), and provide optimal global access services~\cite{TMC_satellite_1}.
Additionally, in the dynamic environments where the coverages of specific areas are constantly changing due to the CubeSat's high orbital speed, it is important to schedule each GS to connect to the CubeSat in order to improve access availability and energy efficiency.
Furthermore, according to the fact that the mobile access demands and requirements of individual GSs are all different depending on their locations, differentiated scheduling algorithms that can take of the characteristics, demands, and requirements of individual GSs are essentially required.

Even though CubeSats can be widely used for next-generation global SAGIN mobile access, CubeSats encounter constraints in delivering global access autonomously, owing to their restricted scales and energy capacities~\cite{zhuhan_uavsatell}.
Hence, despite the capacity of multiple CubeSats to collectively cover extensive areas, there might persist coverage gaps in remote areas, polar regions, or the areas experiencing significant communication burdens.
Moreover, the rapid orbital velocity of CubeSats, approximately $7.5$\,km/s, results in frequent handovers~\cite{TMC_satellite_3}. To maintain uninterrupted global access, it becomes necessary to integrate new \textit{aerial networks} that focus on specific local regions and CubeSats must be considered together~\cite{TMC_UAV_2}.
Finally, despite CubeSats experiencing reduced delay time compared to GEO satellites, their delay time is still significant challenge when contrasted with terrestrial networks (TNs). Consequently, the deployment of innovative NTN devices to support CubeSats is essential for ensuring seamless global access.

To address these challenges, this paper proposes \textit{cooperative and differentiated global SAGIN mobile access involving both CubeSats and aerial networks}.
The aerial networks, possessing enhanced mobility compared to CubeSats that follow predetermined orbits, are capable of more adaptable responses to changing environmental conditions. Consequently, unmanned aerial vehicles (UAVs) are particularly beneficial for establishing networks across diverse regions characterized by uncertainty~\cite{TMC_UAV_3}.
Despite their utility, rotorcrafts consume a significant amount of energy, posing challenges to the seamless global SAGIN mobile access. Therefore, the system discussed in this paper employs high-altitude long-endurance (HALE)-UAVs, which are fixed-wing aircraft, to overcome these limitations.
The HALE-UAVs are distinguished by their capacity for long-distance flights, attributed to their substantial endurance and energy levels.
Furthermore, the attributes of the HALE-UAV, one of fixed-wing aircrafts, enable them to sustain flight longer than rotary-wing aircrafts even in the scenarios where its control systems can be damaged~\cite{intro_8_haleuav}.
Ultimately, HALE-UAVs can supplement CubeSats in providing flexible and extensible coverages for particular regions, such as polar areas lacking signal availability, or the regions burdened with communication overheads~\cite{TMC_UAV_1, TMC_satellite_5}.
Based on these issues and architecture characteristics, we need to design a new global SAGIN scheduling algorithm. 

Moreover, the need for effective scheduling becomes paramount in the scenarios populated by numerous CubeSats and HALE-UAVs.
In order to realize effective scheduling for CubeSats and HALE-UAVs in terms of access availability and energy efficiency, cooperative and differentiated global SAGIN mobile access should be proposed. In this scheduling problem, the goal is to simultaneously improve access availability in terms of QoS and capacity as well as energy efficiency in NTN devices, i.e., CubeSats and HALE-UAVs. To achieve this, we have to consider the hardware restrictions of CubeSats and HALE-UAVs at the same time.
For CubeSats, their geographical coordinates in terms of latitude and longitude as well as the direction vector toward the sun for solar charging undergo real-time alterations due to their orbital movement. 
Furthermore, CubeSats frequently sustain damage from cosmic rays and solar winds. 
Similarly, the flight environment for HALE-UAVs is characterized by dynamic and uncertain conditions, including the presence of vortices and gusts.
Moreover, due to the limited energy levels and capacities of NTN devices, collaboration among these NTN devices is crucial for the simultaneous optimization of energy efficiency and channel capacity.

Distinct from conventional scheduling algorithms, reinforcement learning (RL) exhibits robust performance in \textit{dynamic} and \textit{uncertain} environments~\cite{TMC_RL_1, TMC_RL_2, TVT_GSKIM_Quantum_Rocket}.
MARL proves particularly effective in situations that require \textit{cooperation} among \textit{multiple} NTN devices~\cite{twc_marl}.
Consequently, within global SAGIN mobile access that utilizes CubeSats and HALE-UAVs, MARL-based algorithms based on MARL may be employed, with multiple GSs acting as agents.
Nevertheless, conventional MARL-based schedulers are unable to ensure reward convergence as the \textit{number of agents} and \textit{action dimensions} of GS expands.
To tackle these issues, this paper proposes a novel cooperative and differentiated scheduling algorithm for access availability and energy efficiency in global SAGIN mobile access, leading to the development of \em{quantum MARL (QMARL)}~\cite{tmc24qrl}.
This innovation utilizes the basis measurements, known as \em{projection-valued measure (PVM)}, allowing the proposed QMARL-based scheduler to \textit{diminish the action dimension} to a logarithmic scale~\cite{cikm23baek}.
Furthermore, realistic experimental setting is constructed to demonstrate the superiority and real-world relevance of our proposed QMARL-based scheduler.
This includes the use of actual CubeSat orbital data, aerodynamic information about real HALE-UAVs environments with significant vortices, and the considerations for photovoltaic (PV) charging based on the CubeSats' relative positions to the sun, i.e., the sun side and dark side.
Additionally, each GS, which is an agent, has its own differentiated maximum required channel capacity depending on the region where each GS is located, the population of that region, and the degree of communication overload. Without these settings, excessive global SAGIN mobile access may be provided to GSs that do not require communication services beyond a certain requirement, and GSs with severe communication overload may not be provided with the desired level of global access.
Eventually, this can result in the energy of NTN devices (i.e., CubeSats and HALE-UAVs) being wasted, uselessly.
In conclusion, the efficacy of our proposed QMARL-based scheduler is validated within realistic environments, evidencing that the algorithm fulfills its objectives by simultaneously optimizing the \textit{access availability in SAGIN} and the \textit{energy efficiency in NTN devices} amidst scenarios characterized by \textit{high action dimensions}. Ultimately, in this paper, our considering SAGIN mobile access network is implemented using \textit{multiple GSs, CubeSats, and HALE-UAVs} through our proposed \textit{QMARL-based scheduler} at high action dimensions, and the proposed algorithm is tested in realistic environments to increase real-world applicability.

The main contributions are as follows.
\begin{itemize}
    \item First of all, this paper is the \em{first attempt to employ a QMARL-based global SAGIN mobile access scheduler} for the coordination of CubeSats and HALE-UAVs. The uniqueness of this scheduler stems from its emphasis on \em{reducing the action dimensions through the PVM}. 
    Furthermore, a new reward function is designed and implemented to encourage cooperative global SAGIN mobile access, and efficient and equitable energy usage of NTN devices in multi-CubeSats and multi-HALE-UAVs environments.
    \item Moreover, the proposed QMARL-based scheduler is designed for the coordinated and differentiated global SAGIN mobile access with multiple GSs, CubeSats, and HALE-UAVs. Furthermore, our proposed scheduling also works for energy efficiency in CubeSats and HALE-UAVs. In order to realize this, the reward function of our proposed QMARL-based scheduler is formulated, and thus, it addresses the energy utilization efficiency of CubeSats, taking into account their exposure to the sun side or dark side, which is crucial given their limited energy capacities due to their compact sizes.
    \item Lastly, the efficacy of the proposed algorithm is assessed under realistic experimental environments involving CubeSat that orbits in real space areas as well as HALE-UAV that flies in the real sky. The orbital elements for CubeSats are derived from the \textit{two line element (TLE)}, which provide the foundational data related orbit for these CubeSats. The experiment incorporates a range of \textit{realistic aerodynamic characteristics of HALE-UAVs} to enhance the algorithm's real-world applicability.
    In addition, specific considerations on the differentiated maximum channel capacity in individual GSs show \em{realistic experimental environments depending on the regions where individual GSs are located, the populations of the regions, and the degrees of communication overloads}.
\end{itemize}

The rest of this paper is organized as follows. 
Sec.~\ref{sec:2_preliminaries} presents preliminary knowledge including related work and QMARL.
Sec.~\ref{sec:3_algorithm design} describes the fundamental modeling and 
Sec.~\ref{sec:4_QNN and QMARL} presents the details of our proposed QMARL-based scheduler. 
Sec.~\ref{sec:5_Performance Evaluation} evaluates the performance in realistic environments,
and lastly, Sec.~\ref{sec:6_Concluding Remarks} concludes this paper.

\section{Preliminaries}\label{sec:2_preliminaries}

\subsection{Related Work}
Numerous projects focus on establishing wireless connections to create aerial NTN devices, including UAVs or satellites~\cite{tvt_intro}. 
Given that these rely on battery-based energy management, minimizing energy consumption is crucial to stable operation in unknown environments for the efficient operation of multiple UAVs and satellites~\cite{tvt_multiuav}.
In the literature, the efficient operation of multiple UAVs has garnered significant attention~\cite{zhuhan_gs}. 
Minimizing energy consumption is important to stable operation in unfamiliar environments, necessitating efficient communications~\cite{Related3_plus}. At the same time, efficient scheduling among satellites is imperative to ensure swift responses to diverse sightings and unforeseen events~\cite{sat_scheduling}.
UAVs, characterized by remarkable acquisition flexibility and very high spatial resolution (VHSR), and LEO satellites, capable of providing time-series data across extensive areas, have traditionally been employed independently.
However, the proposed algorithm in \cite{Related7} can minimize total energy costs and reduce time complexity which is crucial for optimizing their effective operation for both UAVs and satellites. Therefore UAVs and satellites must be controlled cooperatively to improve performance~\cite{jsac_together}.
To efficiently manage both UAVs and satellites, numerous studies have demonstrated different methodologies for applying RL algorithms~\cite{related_drl}.
The proposed algorithm in \cite{aimlab_marl} proves the superiority of RL, particularly beneficial in the management of multiple agents. However, to build global SAGIN mobile access, more agents need to be controlled~\cite{related_large}. 
Notably, quantum algorithms have advantages in managing large-scale scenarios, such as those encountered in aerial networks~\cite{aimlab_quantum}.
This paper demonstrates the superiority of using QRL over RL in multi-agent scheduling.

\subsection{Quantum Neural Network}

\begin{figure*}
  \centering
  \subfigure[Qubit on the Bloch sphere.]{
  \includegraphics[width=0.22\linewidth]{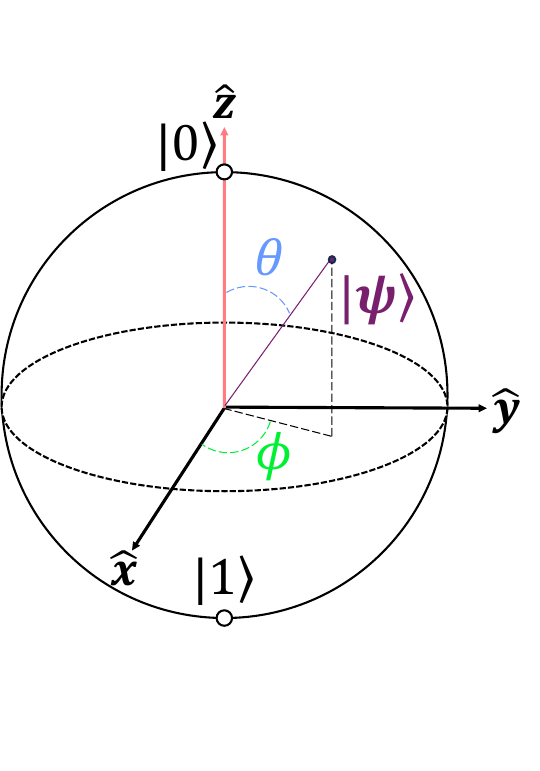}
  \label{fig:Qubit}
  }
  \subfigure[QNN architecture.]{
  \includegraphics[width=0.73\linewidth]{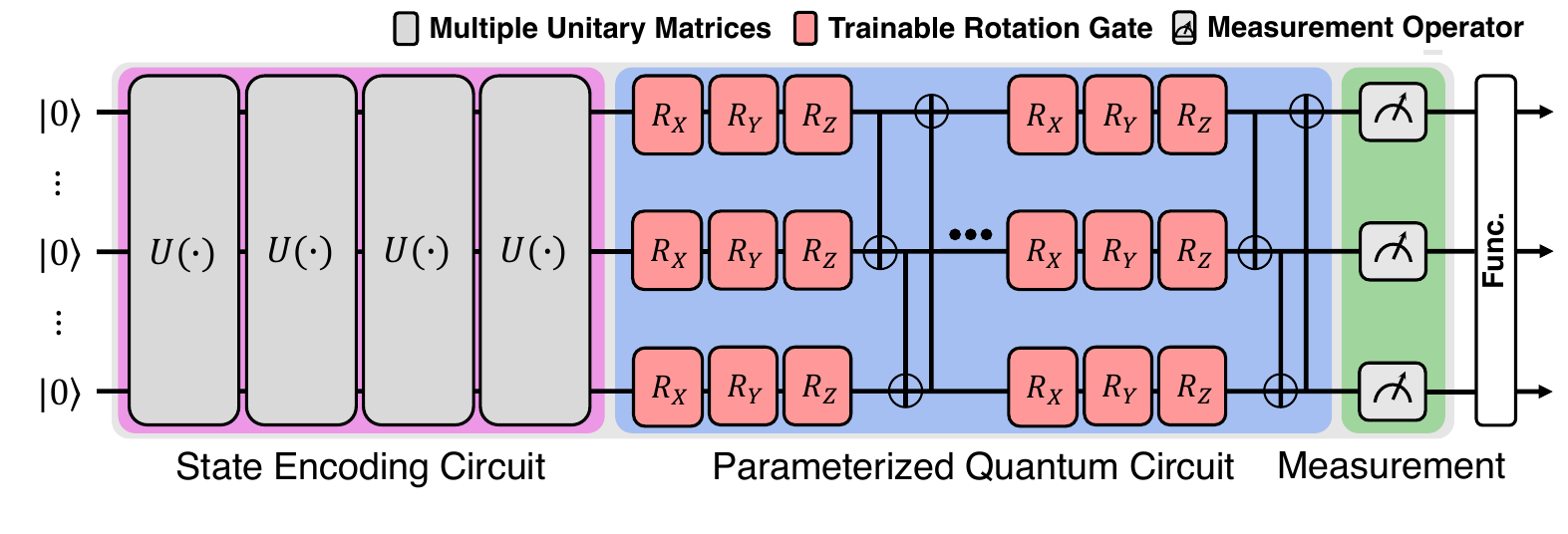}
  \label{fig:QNN}
  }
  \caption{
  Qubit and QNN architecture.
  } 
\end{figure*}

In QNN architectures, a significant deviation from classical neural networks is the utilization of qubits as the unit for basic learning computations~\cite{DBLP:journals/ftsig/Simeone22}. Within quantum systems, qubits stand as the fundamental units of information, and their representation is grounded in the base states of $\ket{0}:=[1,0]^T$ and $\ket{1}:=[0,1]^T$. The representation of a single qubit state can be realized through a normalized 2D complex vector as $\left| \psi  \right>=\alpha \left| 0  \right>+\beta \left| 1  \right>$ and $\left\| \alpha \right\|^{2}+\left\| \beta \right\|^{2}=1$ holds,
where $\left\| \alpha \right\|^{2}$ and $\left\| \beta \right\|^{2}$ denote the probabilities of observing $\left| 0 \right>$ and $\left| 1 \right>$, respectively. 
The QNN computation is carried out over the 3D Bloch sphere, defined as the Hilbert space which represents the quantum domain. Expressing this within the Bloch sphere, which serves as a representation of the quantum domain, it can be geometrically denoted as,  
    $\left| \psi  \right>=\cos(\theta)\left| 0  \right>+e^{i\phi }\sin(\theta) \left| 1  \right>$, where $\theta$ denotes a parameter that determines the probabilities of measuring $\left| 0 \right>$ and $\left| 1 \right>$, and $\phi$ represents the relative phase, respectively, where $0\leq \theta \leq \pi$ and $0\leq \phi < 2\pi$~\cite{DBLP:journals/ftsig/Simeone22}. Fig.~\ref{fig:Qubit} shows a qubit represented over the Bloch sphere. When considering a $q$ qubit system, the representation of quantum states within the system's Hilbert space is as $\ket{\psi} = \sum^{2^q-1}_{l=0} \omega_{l} \ket{l}$, where $\ket{\psi}$ denotes the quantum state, $\ket{l}$ represents $l$-th basis, and $\omega_{l}$ stands for the probability amplitude of $q$ qubit system, respectively. Then, the probability amplitude fulfills $\sum^{2^q-1}_{l=0} |\omega_l|^2 = 1$. A significant component in classical neural networks is a hidden layer, capable of representing linear and nonlinear transformations to achieve accurate function approximation within the neural network. Hence, the primary design consideration factors in QNN involve designing and implementing linear and nonlinear transformations over the 3D sphere. This QNN design facilitates the fundamental enablement of QRL-based control, achieved by incorporating the states and actions of RL-based control as inputs and outputs within QNN architectures.

In QNN architecture, there are three primary components: (i) state encoding, (ii) parameterized quantum circuit (PQC), and (iii) measurement, as illustrated in Fig.~\ref{fig:QNN}.

\begin{itemize}
    \item \BfPara{State Encoding}
The encoder performs the function of converting the classical data, represented as $\zeta_t$ at a specific time $t$, to the initialized quantum state $\ket{0}$. The encoder carries out this function due to the inability of quantum circuits to directly accept classical bits. Through the application of multiple unitary matrices, denoted as $U(\cdot)$, this encoding transformation is achieved mathematically. An important point to highlight is that the encoder does not include any trainable parameters. Thus, the encoded quantum state of the QNN at a specific time $t$ is defined as $\ket{\psi_{0; t}}  =  U_{\text{ENC}}(\zeta_t)\ket{0}^{\otimes q}$, where  the classical data $\zeta_t$ serves as rotation angles within the set of encoding gates $U$.
    \item \BfPara{PQC}
The operations performed by PQC are analogous to the multiplications seen in the accumulated hidden layers of classical neural networks.
Quantum gates can transform the state of qubits through the operations they perform~\cite{DBLP:journals/ftsig/Simeone22}. Within this paper, the following three gates will be introduced: Pauli, \textit{Controlled}, and rotation gates~\cite{DBLP:journals/ftsig/Simeone22}. Outlined below are the definitions for Pauli-$\Gamma$ gates and \textit{Controlled-}$\Gamma$ gates, i.e., 
    $X\!\!=\!\!
    \begin{bmatrix}
        0 & 1\\
        1 & 0
    \end{bmatrix}\!$, 
    $\!Y\!\!=\!\!
    \begin{bmatrix}
        0 &\!-\! i\\
        i & 0
    \end{bmatrix}\!$, 
    $\!Z\!\!=\!\!
    \begin{bmatrix}
        1 & 0\\
        0 & \!-\!1
    \end{bmatrix}\!$, 
    and $C\Gamma \!\!=\!\!
    \begin{bmatrix}
        {\textbf{\textrm{I}}} & 0\\
        0 & \Gamma
    \end{bmatrix}$,
where $i = \sqrt{(-1)}$, $\forall \Gamma \in \left\{ X, Y, Z\right\}$, and ${\textbf{\textrm{I}}}$ stands for the identity matrix, respectively. The Pauli-$\Gamma$ gates perform $180\,^{\circ}$ rotations of the quantum state in the x, y, and z axes of the Bloch sphere. Between two qubits, the \textit{Controlled-}$\Gamma$ gates produce entanglement. Within QNN, rotation gates $R_\Gamma$ featuring the trainable parameters $\theta_k$, defined within the range $[0, 2\pi]$, find widespread utilization. This can be represented as follows: $R_\Gamma(\theta_k)=e^{-i\frac{\theta_k}{2}\Gamma}$. 
Achieving rotations and entanglement of all qubits involves utilizing Pauli-$\Gamma$, \textit{Controlled-}$\Gamma$, and rotation gates. 
At this moment, Pauli-$\Gamma$ gates and $R_\Gamma$ are employed for implementing linear transformations, while the \textit{Controlled-}$\Gamma$ gates are utilized for nonlinear transformations. Therefore, PQC achieves two transformations on the 3D sphere. Consequently, in PQC, it can vary depending on the configuration of the $R_\Gamma$ and \textit{Controlled-}$\Gamma$ gates, and is an important factor in building a QNN. To thoroughly explore trainable rotation parameters and entanglement, we implement multiple quantum layers in this paper, each consisting of $R_\Gamma$ gates within PQC of each QNN. At a specific time $t$, the quantum state of the QNN, denoted as $\ket{\psi_t}$, can be represented as  
    $\ket{\psi_t} = \prod_{l=1}^{L}\nolimits \boldsymbol{U}_l(\theta_t)U_{\text{ENC}}(\zeta_t)\ket{0}^{\otimes q}$,
where $\boldsymbol{U}_l(\theta_t)$ stands for the $l$-th quantum layer at the specific time $t$ with its corresponding set of trainable parameters. Observe that $\boldsymbol{U}_l(\theta_t)$ takes the trainable parameters as inputs, therefore it works differently from the encoder's gates.

\item \BfPara{Measurement}
The quantum state that is acquired by PQC is utilized as the input for measurement.
In this process, quantum data is decoded back to the original format before performing measurements on the input.
The z-axis is commonly used for measurements, but axes in other directions can also be used if they are appropriately defined.
The quantum state collapses and its properties become \textit{observable} after the quantum state is measured.
Upon completion of the decoding procedure, the \textit{observable} property is employed to minimize the loss function. Achieving the expected decoded value of the quantum state $\ket{\psi_t}$ can be accomplished through $\bra{\psi_t}O\ket{\psi_t}$, where $\ket{\psi_t} = \prod_{l=1}^{L} \boldsymbol{U}_l(\theta_t)U_{\text{ENC}}(\zeta_t)\ket{0}^{\otimes q}$, $\bra{\psi_t}$ denotes the conjugate transpose of $\ket{\psi_t}$, and $O$ represents the \textit{observable}, respectively.
\end{itemize}

\subsection{QMARL for Scheduling}
This section investigates the use of QMARL for scheduling CubeSats and HALE-UAVs, presenting a strong argument for its preference over conventional MARL approaches. Conventional MARL has been effective for optimizing decisions in scenarios with relatively small action dimensions. Nonetheless, within intricate systems like integrated networks using CubeSats/HALE-UAVs, characterized by exponentially vast action dimensions, the efficacy of conventional MARL diminishes due to \textit{computational burden} and the \textit{inefficacy} in managing extensive action spaces.
The expansion of the action dimension introduces the challenge of the \textit{curse of dimensionality}~\cite{prel_curse}, a significant impediment in conventional MARL frameworks. QMARL, empowered by quantum computing features such as superposition and entanglement, offers a significant computational edge~\cite{cm2023park}. This quantum advantage allows QMARL to efficiently process large-scale data and complex decision matrices~\cite{aimlab_qmarl_prel_2}, presenting a superior solution for the extensive action dimensions encountered in integrated networks using CubeSats/HALE-UAVs. 
Moreover, the multi-agent dynamics of these integrated networks involving many communicating devices such as multiple GSs, CubeSats, and HALE-UAVs make the scheduling decision-making problem more complex.
QMARL signifies a crucial advancement in overcoming the challenges of high-dimensional and complex scheduling tasks for integrated networks using CubeSats/HALE-UAVs. Its enhanced computational strength and ability to effectively manage multi-agent scenarios establish it as a powerful and efficient approach, facilitating the development of more sophisticated, effective, and dependable SAGIN. 

\section{Modeling}\label{sec:3_algorithm design}

\subsection{Global SAGIN Access Scheduling Modeling}
The considered global SAGIN is illustrated in Fig.~\ref{fig:Overview of this paper} and structured around three principal elements, $N$ GSs, a fleet of $M$ CubeSats, and a group of $L$ HALE-UAVs. Each GS is denoted as $G_i$, $i\in \mathcal{N}$, and note that $|\mathcal{N}|\triangleq N$. In addition, CubeSats and HALE-UAVs are denoted as $S_j$ and $A_l$, respectively, where $S_{j}, j\in\mathcal{M}$ and $A_l, l \in\mathcal{L}$, and also note that $|\mathcal{M}|\triangleq M$ and $|\mathcal{L}|\triangleq L$. 
Our proposed scheduling works by each GS $G_i$ to establish the communications with CubeSats $S^i_j$ or HALE-UAVs $A^i_l$ that are located within the coverage of $G_i$, for network access services. 
The main purpose of this scheduling is for maximizing (i) the residual energy amounts of NTN devices, (ii) the fair energy consumption among NTN devices, and (iii) the global access performance in terms of capacity and QoS, in SAGIN systems.

\subsection{HALE-UAV}
\begin{figure*}[t!]
    \centering
    \subfigure[$z$-axis transformation (yawing).]{
    \includegraphics[width=0.27\linewidth]{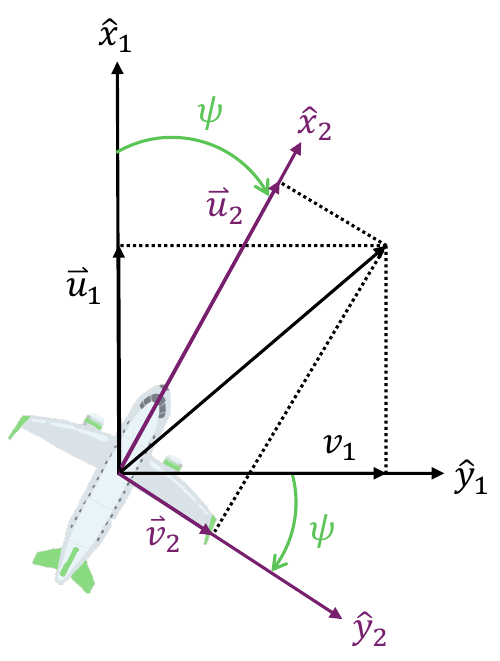}
    \label{fig:yawing}
    }
    \subfigure[$y$-axis transformation (pitching).]{
    \includegraphics[width=0.3\linewidth]{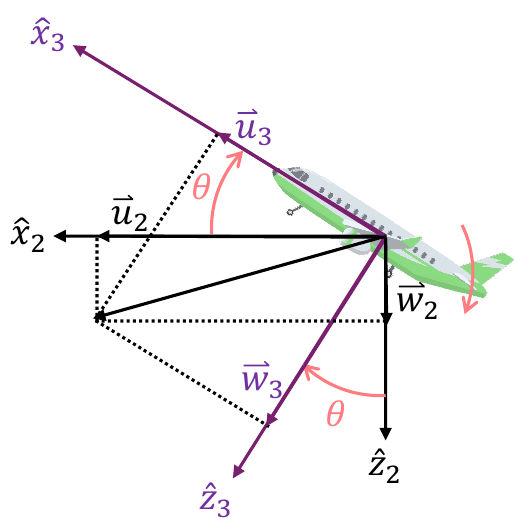}
    \label{fig:pitching}
    }
    \subfigure[$x$-axis transformation (rolling).]{
    \includegraphics[width=0.35\linewidth]{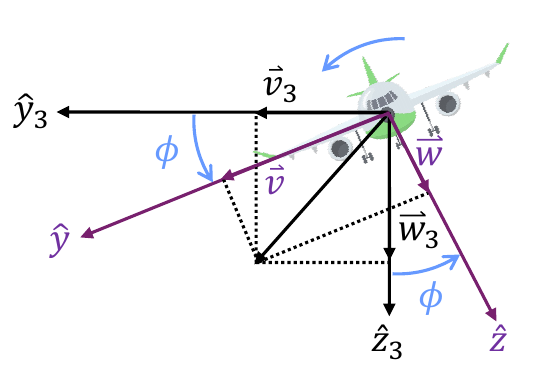}
    \label{fig:Rolling}
    }
\caption{Flight aerodynamics of HALE-UAV.}
\label{fig:Flight Dynamic of aircraft}
\end{figure*}
In order to ensure the maneuvers of HALE-UAVs while maintaining the equilibrium among the energy levels of HALE-UAVs, energy expenditure modeling for HALE-UAV is essential. The required energy is the minimum energy amount to overcome aerodynamic drag and advance in each HALE-UAV. 
The energy is equivalent to the work per unit over time under the force applied to the dynamic system, and it is defined as the dot product of force and velocity. Therefore, the required energy of the $l$-th HALE-UAV at time $t$, denoted as $E^A_l(t)$, is defined as $E^A_l(t)=DV$, where $D$ and $V$ denote its drag and velocity at time $t$, respectively. Here, drag $D$ can be obtained as $D=\frac{1}{2}\rho V^2SC_D=qSC_D$, where $C_D$ is drag coefficient. Because $C_D$ is expressed as $C_D=C_{D_{0}}+kC_L^2$ and $C_L$ is expressed as $C_L=\frac{W}{\frac{1}{2}\rho V^2S}=\frac{W}{qS}$, the required energy of the $l$-th HALE-UAV at time $t$, i.e., $E^A_l(t)$, is,
\begin{multline}
E^A_l(t)=\underbrace{\frac{1}{2}C_{D_{0}}\rho V^3S}_{\textrm{parasite energy, $P_p$}}+\underbrace{\frac{kW^2}{\frac{1}{2}\rho SV}}_{\textrm{induced energy, $P_i$}} \\ =\underbrace{qSC_{D_{0}}V}_{\textrm{parasite energy, $P_p$}}+\underbrace{\frac{W^2kV}{qS}}_{\textrm{induced energy, $P_i$}},
\label{eq:power_required}
\end{multline}
where $C_{D_{0}}$, $\rho$, $V$, $S$, $k$, $W$, and $q$ are the parasite drag coefficient at zero lift, density of the air, velocity, wing surface area, induced drag coefficient, HALE-UAV weight, and dynamic pressure ($ q=\frac{1}{2}\rho V^2$)~\cite{aerodynamic_1}, respectively. As expressed in \eqref{eq:power_required}, the required energy is composed of the \textit{parasite energy} and \textit{induced energy}~\cite{s_jung_TVT}. Here, the parasite energy arises from parasite drag, encompassing \textit{skin friction drag} (drag that varies with the UAV's surface texture), \textit{form drag} (drag that depends on the HALE-UAV's size, structure, and shape), and \textit{interference drag} (drag generated from the interaction between skin friction and form drag)~\cite{helicopter_dynamics}. In addition, the induced energy originates from the drag produced by generating lift. This type of drag is caused by wingtip vortices, resulting from the differential pressure on the wing's upper and lower surfaces, which in turn creates downwash at the wing's rear. Accordingly, $P_p$ increases with the cube of velocity, whereas $P_i$ is inversely related to velocity, demonstrating the dynamics of aerodynamic drag in relation to the UAV's velocity~\cite{parasite_drag}.

On the other hand, velocity $V$ is computed as the aggregate of velocities along each axis, formulated as $V=\sqrt{u^2+v^2+w^2}$, where $u$, $v$, and $w$ represent the velocities over the $x$-, $y$-, and $z$-axes of body axis coordinate system, respectively.
Here, velocity $V$ in \eqref{eq:power_required} is the velocity based on the body axis coordinate system of aircraft. 
Nevertheless, due to the fact that the velocities of HALE-UAVs for each axis are determined with the relation to the ground coordinate system, it is imperative to utilize coordinate transformation matrices.
Therefore, velocities $u_1$, $v_1$, and $w_1$ in the ground coordinate system are transformed into the velocities $u$, $v$, and $w$ within the body axis coordinate system through multiplication by the coordinate transformation matrices $L_1$, $L_2$, and $L_3$, which is expressed as, 
\begin{equation}
\begin{bmatrix}
u \\
v \\
w
\end{bmatrix}
=L_1 \times L_2 \times L_3 \times
\begin{bmatrix}
u_1 \\
v_1 \\
w_1
\end{bmatrix},
\end{equation}
where $L_1$, $L_2$, and $L_3$ are the transformation matrices over the $z$-axis, $y$-axis, and $x$-axes, sequentially. 
The geometric relationships among these transformations are illustrated in Fig.~\ref{fig:Flight Dynamic of aircraft}, and the transformation of coordinates for each axis can be articulated via,
\begin{equation}
\begin{bmatrix}
u_2 \\
v_2 \\
w_2
\end{bmatrix}
=
\underbrace{\begin{bmatrix}
\cos \psi  & \sin\psi &  0 \\
-\sin \psi & \cos \psi & 0 \\
0 & 0 & 1 \\
\end{bmatrix}}_{L_1}
\begin{bmatrix}
u_1 \\
v_1 \\
w_1
\end{bmatrix}, 
\end{equation}
\begin{equation}
\begin{bmatrix}
u_3 \\
v_3 \\
w_3
\end{bmatrix}
=
\underbrace{\begin{bmatrix}
\cos \theta  & 0 &  -\sin\theta \\
0 & 1 & 0 \\
\sin \theta & 0 & \cos \theta \\
\end{bmatrix}}_{L_2}
\begin{bmatrix}
u_2 \\
v_2 \\
w_2
\end{bmatrix},
\end{equation}
\begin{equation}
\begin{bmatrix}
u \\
v \\
w
\end{bmatrix}
=
\underbrace{\begin{bmatrix}
1  & 0 &  0 \\
0 & \cos\phi & \sin\phi \\
0 & -\sin\phi & \cos \phi \\
\end{bmatrix}}_{L_3}
\begin{bmatrix}
u_3 \\
v_3 \\
w_3
\end{bmatrix},
\end{equation}
where $\psi$, $\theta$, and $\phi$ represent the rotations over the $z$-, $y$-, and $z$-axes, respectively. 
Within the real flight environment of HALE-UAVs, such disturbances are attributable to turbulence and wind gusts, which have the potential to alter the UAV's rotational orientation.
Amidst conditions where turbulence and gusts are prevalent across all axes, the goal of HALE-UAV is to simultaneously optimize the global access performance of the integrated network and the energy use of HALE-UAV.
Details pertaining to the HALE-UAV deployed in this paper are compiled in Table~\ref{tab:parameters of uav}.
\begin{table}[t!]
\centering
\footnotesize
\caption{Specifications of HALE-UAV.}
\renewcommand{\arraystretch}{1.0}
\begin{tabular}{l||r}
\toprule[1pt]
\textsf{\textbf{Notation}} & \textsf{\textbf{Value}} \\ \midrule
Mass of HALE-UAV, $m$ & 1,815\,[$\mathrm{kg}$]\\
Acceleration of gravity, g & 9.81\,[$\mathrm{m/s^2}$]\\
Weight of HALE-UAV, ${W}={mg}$ & 17,799\,[$\mathrm{N}$]\\ 
Wing surface area, S & 6.61\,[$\mathrm{m^{2}}$] \\
Density of the air, $\rho$ & 0.089\,[$\mathrm{kg/m^{3}}$] \\
Parasite drag coefficient at zero lift, $C_{D_0}$ & 0.045 \\
Induced drag coefficient, $k$ & 0.052 \\
\bottomrule[1pt]
\end{tabular}
\label{tab:parameters of uav}
\end{table}

\subsection{CubeSat} \label{sec:Orbital elements of CubeSat}

\subsubsection{Two Line Element (TLE)}
\begin{figure}
  \centering
  \includegraphics[width=1.0\linewidth]{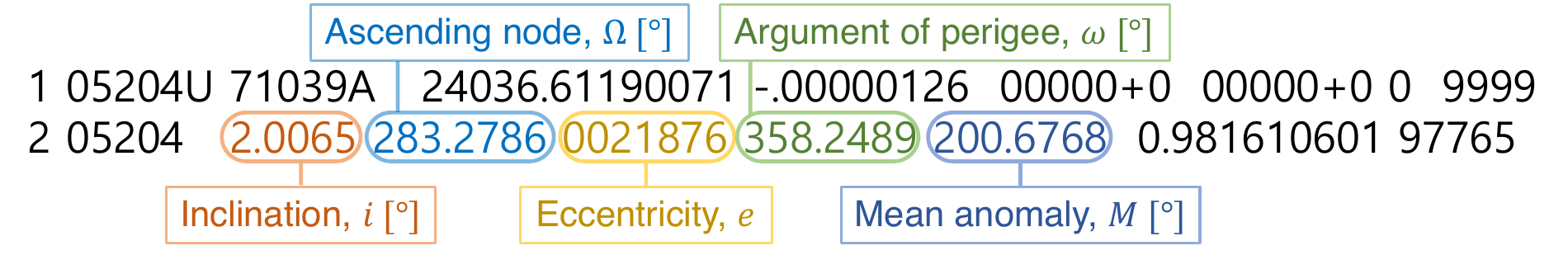}
  \caption{TLE configuration of the satellite used in the experiment..}
  \label{fig:TLE}
\end{figure}
In order to observe the orbital mechanics of CubeSats, TLE is essentially required. Originating from the North American Aerospace Defense Command (NORAD), TLE contains the vital details concerning the trajectories of objects orbiting the Earth, especially for CubeSats. NORAD, tasked with the surveillance and cataloging of space debris, introduced the TLE format to  effectively disseminate orbital information.  
The structure of TLE consists of two lines as illustrated in Fig.~\ref{fig:TLE}, detailing specific orbital parameters and CubeSat characteristics.
Fig.~\ref{fig:TLE} displays the TLE for \textit{OPS-3811}, a CubeSat utilized in the experiment, encompassing orbital elements such as inclination ($i$), ascending node ($\Omega$), eccentricity ($e$), argument of perigee ($\omega$), and mean anomaly ($M$). 
The inclination ($i$) signifies the CubeSat's orbital plane angle relative to the equatorial plane of the Earth. 
The ascending node ($\Omega$) specifies the location where the CubeSat's orbit crosses the equatorial plane from south to north, also known as the right ascension of the line of nodes. 
The eccentricity ($e$) is a measure of how far a CubeSat's elliptical orbit deviates from a circle.
The argument of perigee ($\omega$) is the angle from the line of nodes to the perigee of the orbit.
The mean anomaly ($M$) indicates the CubeSat's current position within its orbit, assuming a circular path with the same semi-major axis ($a$). In other words, the mean anomaly is the angle between the current position of the CubeSat and the perigee of the orbit, assuming that the CubeSat moves at an average speed when moving along an elliptical orbit.
These TLE data, such as $e$ and $\Omega$, are instrumental in calculating the CubeSat's latitude, longitude, facilitating the determination of $x^i_{j}(t)$ between $G_i$ and $S^i_j$, by \eqref{eq:obj-2}.

\subsubsection{Orbital Elements of CubeSats}

\begin{figure*}
  \centering
  \subfigure[Geometry of orbital elements.]{
  \includegraphics[width=0.40\linewidth]{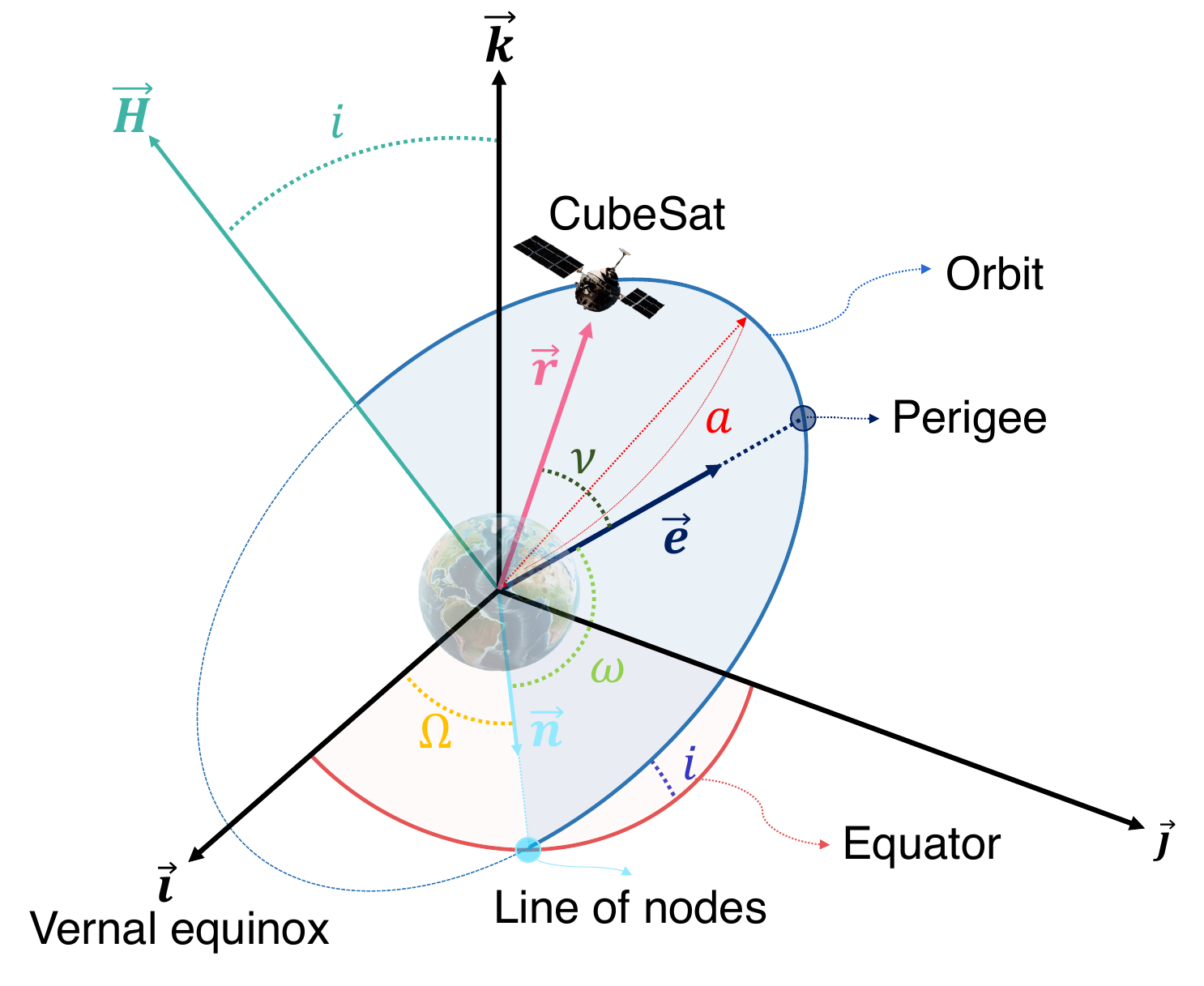}
  \label{fig:Geometry of orbital elements}
  }
  \subfigure[Coordinate of the Earth.]{
  \includegraphics[width=0.45\linewidth]{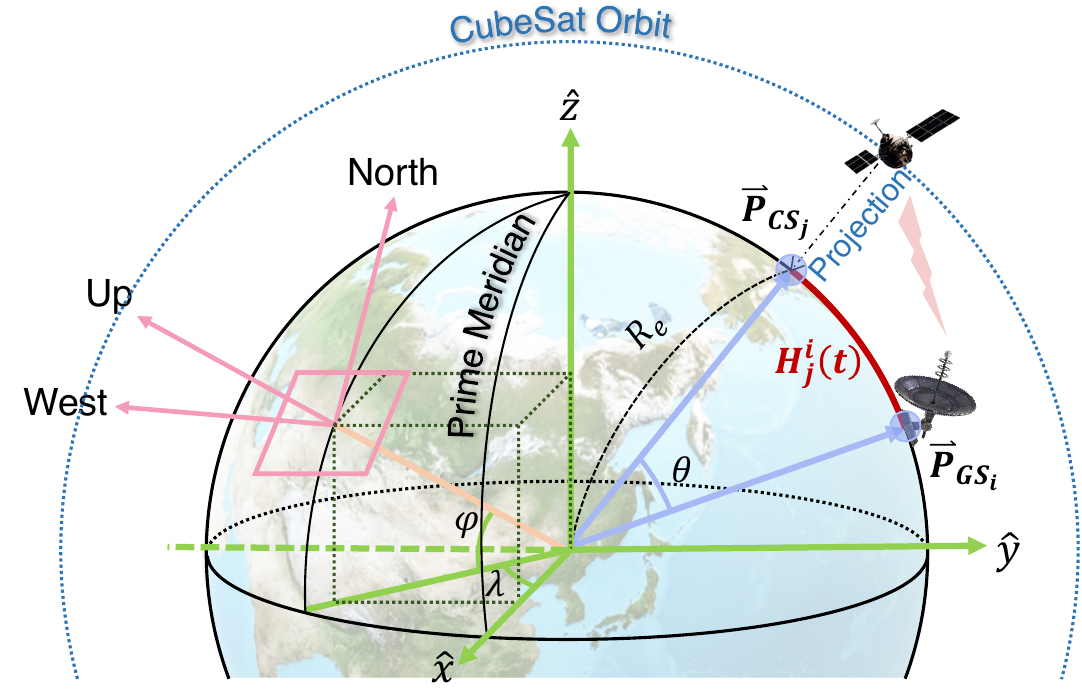}
  \label{fig:Coordinate of the Earth}
  }
  \caption{Orbital elements of CubeSat and the geometric relationship of great circle distance between two CubeSats.}
  \label{fig:Orbital Elements}
\end{figure*}

As mentioned, the orbital elements expressed in TLE include eccentricity ($e$), inclination ($i$), right ascension of the ascending node ($\Omega$), argument of perigee ($\omega$), and mean anomaly ($M$). The orbital elements that are not in TLE, such as semi-major axis ($a$), eccentric anomaly ($E$), and true anomaly ($\nu$), are obtained using the orbital elements in TLE.
Fig.~\ref{fig:Orbital Elements}(a) presents the geometric representation of orbital elements. The semi-major axis ($a$), illustrated with a green line, denotes the CubeSat's orbit's longest radius, crucial for calculating its eccentricity ($e$). The eccentricity itself measures how much the orbit deviates from a perfect circle, with values close to $0$ indicating near circularity and values near $1$ highlighting an elliptical shape. 
The eccentricity vector ($\overrightarrow{e}$) is a vector that goes from the center of the CubeSat's orbit to the perigee of the orbit.
Additionally, the orbital inclination ($i$) is assessed as the angle between the orbit's normal axis ($\overrightarrow{k}$) and its angular momentum vector ($\overrightarrow{H}$), with the latter perpendicular to the plane of the orbit, thereby quantifying the orbit's tilt with respect to the equatorial plane of the Earth. 
The ascending node ($\Omega$) signifies the \textit{line of nodes}'s longitude, which is the point where the CubeSat's orbital plane intersects the Earth's equatorial plane.
The argument of perigee ($\omega$) is defined by the angle from the ascending node vector 
($\overrightarrow{n}$) to the eccentricity vector ($\overrightarrow{e}$), with $\overrightarrow{n}$ directing towards the \textit{line of nodes}, depicted as a \textit{sky blue line} in Fig.~\ref{fig:Geometry of orbital elements}. This angle delineates the orbit's orientation relative to the equator, marking the perigee's location.
The mean anomaly ($M$) is a parameter for predicting the position of a CubeSat moving along an elliptical orbit over time, and is expressed as an angle representing the average position of the object within the orbital period,
aiding in the calculation of the eccentric anomaly ($E$). 
In an elliptical orbit, the CubeSat's velocity changes as it passes through periapsis (the closest point) and apogee (the farthest point), but mean anomaly does not take these velocity changes into account and assumes that it moves at a uniform velocity.
Therefore, a difference may occur between the actual position of the CubeSat and the position calculated by mean anomaly, and eccentric anomaly and true anomaly are used to correct this difference.
The mean anomaly does not directly correspond to the actual CubeSat position, but is used as an initial value to calculate more accurate positions, such as the eccentric anomaly and true anomaly, using the eccentricity of the orbit and other orbital elements. Therefore, the mean anomaly plays an important role when modeling trajectories as a \textit{function of time}.
Finally, the true anomaly ($\nu$) is the angle from the perigee to the CubeSat's actual position, represented by the angle between vectors $\overrightarrow{r}$ and $\overrightarrow{e}$, where $\overrightarrow{r}$ points from the origin of the coordinate system to the CubeSat, and the coordinate axis $\overrightarrow{i}$ aims towards the vernal equinox.

\subsubsection{Latitude and Longitude of CubeSat}

To ascertain the locations of CubeSats change over time, their positions are represented through coordinates of \textit{latitude} ($p^\phi_j(t)$) and \textit{longitude} ($p^\lambda_j(t)$) within the orbital coordinate systems.
Given that the CubeSat's unprocessed data in TLE consist of the coordinates in the \textit{celestial coordinate systems}, the transformation to the \textit{orbital coordinate systems} is required for the derivation of latitude and longitude.
The latitude and longitude that change over time for each CubeSat are calculated through TLE, which is raw CubeSat data.
Consequently, the latitude ($p^\phi_j(t)$) and longitude ($p^\lambda_j(t)$) pertaining to the current position of CubeSat $S^i_j$, i.e., the $j$-th CubeSat located within the coverage of the $i$-th GS, are articulated as, 
$p^\phi_j(t) = \sin^{-1}\left(\frac{R_{f}[3]}{\lVert R_f \rVert}\right)$
and
$p^\lambda_j(t) = \cos^{-1}\left(\frac{R_{f}[1]}{\lVert R_f \rVert \cos\phi}\right)$,
where $R_{f}[1]$ and $R_{f}[3]$ refer to $R_f$'s first and third elements, and this $R_{f}$ is defined as,
\begin{equation}
R_f\triangleq [C_1\times C_2\times C_3\times C_4]\times V_4.
\label{eq:R_f}
\end{equation}
In \eqref{eq:R_f}, the coordinate transformation matrices, $C_1$, $C_2$, $C_3$, and $C_4$, are 
\begin{align}\footnotesize 
   & C_1 = \begin{bmatrix}
    \cos(\Omega) & \sin(\Omega) & 0 \\
    -\sin(\Omega) & \cos(\Omega) & 0 \\
    0 & 0 & 1
\end{bmatrix}, 
   C_2 = \begin{bmatrix}
    1 & 0 & 0 \\
    0 & \cos(i) & \sin(i) \\
    0 & -\sin(i) & \cos(i)
\end{bmatrix}, \nonumber \\
\footnotesize    
   & C_3 = \begin{bmatrix}
    \cos({\omega}) & \sin({\omega}) & 0 \\
    -\sin({\omega}) & \cos({\omega}) & 0 \\
    0 & 0 & 1
\end{bmatrix}, 
   C_4 = \begin{bmatrix}
    \cos({\theta}) & \sin({\theta}) & 0 \\
    -\sin({\theta}) & \cos({\theta}) & 0 \\
    0 & 0 & 1
\end{bmatrix}, \nonumber
\end{align}
where $\theta$ is the angle by which the Earth has rotated in $t$.
Therefore, $\theta$ represents the product of the Earth's rotational angular velocity and the time interval $t$.
Lastly, $V_4$ in \eqref{eq:R_f} is,
\begin{equation} 
V_4= 
\begin{bmatrix}
    r\cos({\nu})~~
    r\sin(\nu)~~
    0
\end{bmatrix}^{T},
\label{eq:V_4}
\end{equation}
where $r$ denotes the conic section, and this $r$ is a clue to compute the distance between the center of the elliptical orbit and CubeSat. Additionally, $\overrightarrow{r}$ is the vector pointing from the center of the elliptical orbit to the current position of CubeSat.
Therefore, the current coordinates of CubeSat measured in the celestial coordinate system are expressed as \eqref{eq:V_4}. However, in order to calculate the CubeSat's latitude and longitude that change over time, $V_4$ in the celestial coordinate system must be converted to the orbital coordinate system, and the previously defined coordinate transformation matrices are utilized. The corresponding coordinate transformation matrices, denoted as $C_1$, $C_2$, $C_3$, and $C_4$, facilitate the conversion of celestial coordinate systems into orbital coordinate systems.
Finally, $r$ in \eqref{eq:V_4} is determined by
\begin{equation} 
r=\frac{H^2/\mu }{1+e\cos(\nu)}, \label{eq:conic section}
\end{equation}
where $\mu$ and $H$ represents the standard gravitational parameter and angular momentum, respectively, where 
$H \triangleq \sqrt{\mu a(1-e^2)}$   
and 
$\nu=2\tan^{-1}\left(\sqrt{\frac{1+e}{1-e}}\tan\left(\frac{E}{2}\right)\right)$, 
where 
$E=M+e\sin{M}$.
Here, the data from TLE are transformed into geographical coordinates, i.e., latitude and longitude, over time.
The constants needed to calculate the latitude and longitude of a CubeSat that change over time through TLE are summarized in Table~\ref{tab:Constant or orbit}.

\begin{table}[t!]
\centering
\footnotesize
\caption{Parameter Settings for CubeSat Position Calculations}
\renewcommand{\arraystretch}{1.0}
\begin{tabular}{l||r}
\toprule[1pt]
\textbf{Constant} & \textbf{Value} \\ \midrule
Gravitational Constant, $G$ & 6.673 $e$-20 \\
Mass of the Earth, $M_e$ & 5.974 $e$+24 kg \\
Radius of the Earth, $R_e$ & 6.378~e+6 m \\
Standard Gravitational Parameter, $\mu$ = $GM_e$ &  3.986~e+14 $m^3$ $s^{-2}$ \\
\bottomrule[1pt]
\end{tabular}
\label{tab:Constant or orbit}
\end{table}

\subsubsection{Distance between GS and CubeSat}
The distance between GSs and NTN devices (i.e., CubeSats and HALE-UAVs) can be formulated as follows.

\begin{lemma}\label{lem:1}
The distance between $G_i$ and $S^i_j$, varies over time due to the updated latitude and longitude of the CubeSat. It can be formulated as, 
\begin{equation}
d^i_j(t) = \sqrt{H^i_j(t)^2+V^i_j(t)^2},
\label{eq:distance}
\end{equation}
where $H^i_j(t)$ and $V^i_j(t)$ represent the respective horizontal and vertical distances between $G_i$ and $S_j^i$, and note that $V^i_j(t)$ indicates the altitude of $S^i_j$ relative to $G_i$.
Then,  
\begin{multline}
H^i_j(t)=R_{e}\cos^{-1}(\cos p^\phi_i(t) \cos p^\phi_j(t)\cos(p^\lambda_i(t)-p^\lambda_j(t)) \\ +\sin p^\phi_i(t)\sin p^\phi_j(t)),
\label{eq:distance_H}
\end{multline}
where $p^\phi_i(t)$ and $p^\lambda_i(t)$ denote the latitude and longitude of $G_i$; and $R_e $ is the radius of the Earth.
\begin{proof}
As illustrated in Fig.~\ref{fig:Coordinate of the Earth}, $\vec{P}_{GS_{i}}$ and $\vec{P}_{CS_{j}}$ are positioned on the surface of the Earth. These vectors are denoted as $\vec{P}_{GS_{i}}=(x_i,y_i,z_i)$ and $\vec{P}_{CS_{j}}=(x_j,y_j,z_j)$, correspondingly, where $\vec{P}_{GS_{i}}$ and $\vec{P}_{CS_{j}}$ are identified as coordinate vectors along with $x$-, $y$-, and $z$-axes, respectively. In addition, the angular difference between $\vec{P}_{GS_{i}}$ and $\vec{P}_{CS_{j}}$, i.e., $\theta$, can be obtained as,
\begin{multline}
\theta =\cos^{-1}\frac{\vec{P}_{GS_{i}}\cdot \vec{P}_{CS_{j}}}{\left\| \vec{P}_{GS_{i}} \right\| \left\| \vec{P}_{CS_{j}} \right\|} \\ =\cos^{-1}\frac{x_ix_j+y_iy_j+z_iz_j}{\sqrt{x_i^2+y_i^2+z_i^2}\sqrt{x_j^2+y_j^2+z_j^2}},
\end{multline}
where $x_i$, $y_i$, $z_i$, $x_j$, $y_j$, and $z_j$ can be represented as,
\begin{align}
\begin{bmatrix}
x_i \\
y_i \\
z_i
\end{bmatrix}
=
\begin{bmatrix}
R_e\cos p_i^\phi(t) \cos p_i^\lambda(t) \\
R_e\cos p_i^\phi(t) \sin p_i^\lambda(t) \\
R_e\sin p_i^\phi(t)
\end{bmatrix}, 
\\
\begin{bmatrix}
x_j \\
y_j \\
z_j
\end{bmatrix}
=
\begin{bmatrix}
R_e\cos p_j^\phi(t) \cos p_j^\lambda(t) \\
R_e\cos p_j^\phi(t) \sin p_j^\lambda(t) \\
R_e\sin p_j^\phi(t)
\end{bmatrix},
\label{eq:proof_matrix}
\end{align}
where $p_i^\phi(t)$, 
    $p_i^\lambda(t)$,
    $p_j^\phi(t)$, and 
    $p_j^\lambda(t)$ are 
    the latitude of $\vec{P}_{GS_{i}}$, 
    the longitude of $\vec{P}_{GS_{i}}$, 
    the latitude of $\vec{P}_{CS_{j}}$, 
    and
    the longitude of $\vec{P}_{CS_{j}}$, at $t$, respectively.
    Given that the magnitudes of these vectors are equivalent, 
    $\sqrt{x_i^2+y_i^2+z_i^2} = \sqrt{x_j^2+y_j^2+z_j^2} = R_e$, and thus, 
    $x_ix_j+y_iy_j+z_iz_j = R_{e}^2\cos^{-1}(\cos p_i^\phi(t)\cos p_j^\phi(t)\cos(p_{i}^{\lambda}(t)-p_{j}^{\lambda}(t))+ \sin p_{i}^\phi(t) \sin p_{j}^\phi(t))$ by \eqref{eq:proof_matrix}.
    Therefore, according to the fact that $H^i_j(t)$  is derived from $R_e\theta$, which is depicted as the red line in Fig.~\ref{fig:Coordinate of the Earth}, 
        $H^i_j(t)=R_{e}\cos^{-1}(\cos p_{i}^{\phi}(t)\cos p_{j}^{\phi}(t)\cos(p_{i}^{\lambda}(t) - p_{j}^{\lambda}(t))+ \sin p_{i}^{\phi}(t)\sin p_{j}^{\phi}(t))$.
\end{proof}
\end{lemma}

Similarly, the distance between $G_i$ and the $l$-th HALE-UAV within the coverage of $G_i$, i.e., denoted as $A^i_l$, is determined based on the latitude ($p^\phi_l(t)$) and longitude ($p^\lambda_l(t)$) of $A^i_l$, calculated as $d^i_l(t) = \sqrt{H^i_l(t)^2+V^i_l(t)^2}$,
where $H^i_l(t)$ and $V^i_l(t)$ are the horizontal and vertical distances, and note that $V^i_l(t)$ indicates the altitude of $A_l^i$ relative to $G_i$, due to \eqref{eq:distance}. 
Furthermore, according to \eqref{eq:distance_H}, $H^i_l(t)=R_{e}\cos^{-1}(\cos p^\phi_i(t) \cos p^\phi_l(t)\cos(p^\lambda_i(t)-p^\lambda_l(t))+\sin p^\phi_i(t)\sin p^\phi_l(t))$, 
where $p^\phi_l(t)$, and $p^\lambda_l(t)$ denote the latitude and longitude of the $l$-th HALE-UAV at time $t$, respectively.

\section{Problem Formulation and Algorithm Design}\label{sec:4_QNN and QMARL}

\subsection{Main Objective for Global SAGIN Mobile Access}
The purpose of our proposed QMARL-based scheduler in SAGIN is to preserve the residual energy of NTN devices as much as possible while each GS improves the global access performance in terms of access availability and energy efficiency.
Therefore, when each GS schedules CubeSats and HALE-UAVs for global access, it is important to simultaneously optimize the global access performance and the residual energy of NTN devices. To achieve this goal, corresponding reward function should designed for MARL based algorithm design.
The main objective of global SAGIN mobile access for each $i$-th GS can be formulated as,
\begin{equation}
\max_{x^i_{j,l}(t)\in\{0,1\}}: \lim_{\mathcal{T}\rightarrow\infty}
\frac{1}{\mathcal{T}}\sum_{t=0}^{\mathcal{T}-1}\nolimits \sum_{\forall j\in M^i, \forall l\in L^i}\nolimits R_{i}(d^i_{j,l}(t), x^i_{j,l}(t)),
\label{eq:obj-1}
\end{equation}
where $d^i_{j,l}(t)$ and $x^i_{j,l}(t)$ represent the distance and the scheduling vector between $G_i$ and the NTN device within the coverage of $G_i$ (i.e., $S^i_j$ or $A^i_l$) at $t$, respectively. 
In addition, $M^i$ and $L^i$ in \eqref{eq:obj-1} stand for the sets of CubeSats and HALE-UAVs within the coverage of $G_i$. 
Furthermore, $\sum_{\forall j \in M^i, \forall l \in L^i} x^i_{j,l}(t) \leq \bar{H}_i, \forall x^i_{j,l}(t)\in \{0,1\}, \forall j \in M^i, \forall l \in L^i$ holds where $\bar{H}_i$ means the maximal number of acceptable NTN devices ($S^i_j$ or $A^i_l$) that $G_i$ can monitor. 
Lastly, $R_{i}(d^i_{j,l}(t), x^i_{j,l}(t))$ is our utility function for seamless global access, and it can be formulated as,
\begin{equation}
    R_i(d^i_{j,l}(t), x^i_{j,l}(t)) = U_i(d^i_{j,l}(t), x^i_{j,l}(t)) - C_i(d^i_{j,l}(t), x^i_{j,l}(t)), 
\label{eq:obj-2}
\end{equation}
where $U_i(d^i_{j,l}(t), x^i_{j,l}(t))$ and $C_i(d^i_{j,l}(t), x^i_{j,l}(t))$ stand for the utility and cost functions.
In \eqref{eq:obj-2}, 
\begin{equation}
U_i(d^i_{j,l}(t), x^i_{j,l}(t)) = \sum_{\forall j \in M^i, \forall l \in L^i}\nolimits \textbf{q}(d^i_{j,l}(t)) \cdot \xi_{j,l}^{SA}(t) \cdot x^i_{j,l}(t),
\label{eq:function_U}
\end{equation}
where $\textbf{q}(d^i_{j,l}(t))$ and $\xi_{j,l}^{SA}(t)$ denote the quality function and capacity of the link between $G_i$ and its associated NTN device ($S^i_j$ or $A^i_l$). In \eqref{eq:function_U}, the quality function can be generalized as~\cite{ton2005quality},
\begin{equation}
\textbf{q}(d^i_{j,l}(t)) \triangleq \left(1+\exp^{-\xi_{1}\left(\Lambda^i_{j,l}(d^i_{j,l}(t))-\xi_{2}\right)}\right)^{-1}, 
\label{eq:quailty_function}
\end{equation}
where the data rate $\Lambda^i_{j,l}(d^i_{j,l}(t))$ depends on bandwidth ($\mathrm{W}$) and signal-to-noise ratio (SNR), which is denoted as $\Gamma$, thus,
\begin{equation}
\Lambda^i_{j,l}(d^i_{j,l}(t)) = \mathrm{W} \cdot \log_{2}\left(1+\Gamma(d^i_{j,l}(t))\right).
\label{eq:function_rate}
\end{equation}
Additionally, the cost function in \eqref{eq:obj-2} is expressed as,
\begin{multline}
C_i(d^i_{j,l}(t), x^i_{j,l}(t)) = 
\sum_{\forall j \in M^i}\limits E^S_{j}(d^i_{j}(t), x^i_{j}(t)) \cdot \underbrace{\sigma_i^{S}(t)}_{\textrm{(cooperation)}} \\ +
\sum_{\forall l \in L^i}\limits E^A_{l}(d^i_{l}(t), x^i_{l}(t)) \cdot \underbrace{\sigma_i^{A}(t)}_{\textrm{(cooperation)}}, 
\label{eq:obj-6}
\end{multline}
where $E^S_{j}(d^i_{j}(t), x^i_{j}(t))$ and $E^A_{l}(d^i_{l}(t), x^i_{l}(t))$ represent the normalized energy expenditure of $S^i_j$ and $A^i_l$, respectively.
In \eqref{eq:obj-6}, $\sigma_i^{S}(t)$, and $\sigma_i^{A}(t)$ quantify the standard deviation of the residual energy levels for $S^i_j$ and $A^i_l$.
The \textit{cooperation} highlighted in \eqref{eq:obj-6} is essential for reducing the variance of each NTN device (CubeSat or HALE-UAV)'s energy status, thereby it can avert the disproportionate energy usage of any specific CubeSat or HALE-UAV as well as promote collaborative operations for minimizing total energy expenditure.

Furthermore, the total energy expenditure, i.e., $E^S_{j}(d^i_{j}(t), x^i_{j}(t))$ and $E^A_{l}(d^i_{l}(t), x^i_{l}(t))$, corresponds to the amount of energy utilized during communications between $G_i$ and its associated NTN device ($S^i_j$ or $A^i_l$).
The energy consumed in $S^i_j$, i.e., $E^S_{j}(d^i_{j}(t), x^i_{j}(t))$, and also in $A^i_l$, i.e., $E^A_{l}(d^i_{l}(t), x^i_{l}(t))$, are limited by their specific maximum capacities, $\bar{e}_j$ for $S^i_j$ and $\bar{e}_l$ for $A^i_l$, which can be expressed as $E^S_{j}(d^i_{j}(t), x^i_{j}(t)) \leq \bar{e}_{j}, \forall j \in M^i$ and $E^A_{l}(d^i_{l}(t), x^i_{l}(t)) \leq \bar{e}_{l}, \forall l \in L^i$, respectively.
Furthermore, the maximum capacity of $G_i$ is also taken into account, i.e.,
\begin{multline}
\xi^{GS}_{i}(t) + \sum_{\forall j \in M^i}^{}\nolimits \xi^S_j(t)\cdot x^i_j(t) \\ + \sum_{\forall l \in L^i}^{}\nolimits \xi^A_l(t)\cdot x^i_l(t) \leq \bar{\xi}_i =\frac{\varrho }{1+e^{-\zeta (t-\tau )}},
\end{multline}
where $\xi^{GS}_{i}(t)$, $\xi^S_j(t)$, $\xi^A_l(t)$, and $\bar{\xi}_i$, are the capacity of $G_i$, the capacity of $S^i_j$, the capacity of $A^i_l$, and the maximum capacity of the $G_i$, respectively, and the $\bar{\xi}_i$ varies depending on the region where each GS is located, the population of that region, and the degree of communication overloads.
Additionally, $\varrho$, $\zeta$, $t$, and $\tau$ are the maximum of logarithmic quality function curve, control factor the steepness of the curve, time, and midpoint of the curve, respectively.

\subsection{Reinforcement Learning Modeling}
According to the dynamics of CubeSats and HALE-UAVs under uncertain environments, the rapid and unexpected state changes occur over time. These dynamics and uncertain environments are obviously obstacles for large-scale global SAGIN mobile access scheduling, which can be modelled with \textit{combinatorics optimization}. For more details, these scheduling problems are generally formulated as integer programming (IP), which are known for their non-deterministic polynomial (NP)-hard complexity, making them particularly difficult to solve using conventional methods. Therefore, it is highly advantageous to re-formulate the original optimization framework into RL-based sequential discrete-time decision-making for time-average scheduling utility maximization. 
Additionally, in the environment formalized through RL, GS constantly interacts with the environment and learns the optimal policy in the process, therefore RL can be a good solution in such a very dynamic and uncertain environment.
However, to implement realistic global access in SAGIN, many GSs, CubeSats, and HALE-UAVs are needed. Because multiple GSs are required, this changes the form of the problem from RL to MARL scheduling, and because multiple CubeSats and HALE-UAVs must be used, the action dimension of the GS increases exponentially as the number of these NTN devices increases.
The conventional MARL has a fatal problem that as the number of GS increases, or as the number of actions that GS can select, that is, the number of CubeSats and HALE-UAVs increases, GS suffers from the \textit{curse of dimensionality} and its learning performance deteriorates.
This paper undertakes such a re-formulation using \textit{QMARL}, proposing a novel approach for tackling the complexities of scheduling in time-varying dynamic environments.
QMARL utilizes QNN and is free from the curse of dimensionality, which is the big problem in conventional MARL.
If QMARL is used to implement realistic global access in SAGIN, seamless global access can be achieved by simultaneously optimizing global access performance and the residual energy of NTN devices even when using numerous GS, CubeSat, and HALE-UAV.

\BfPara{State}
In our considering aerial network with CubeSats and HALE-UAVs, the \textit{state} is defined by the observational data collected by $G_i$, denoted as $\mathcal{S}_i(t)$, and it can be as follows,
\begin{multline}
\mathcal{S}_i(t) \triangleq \{P_i(t), \xi_i(t),
\bigcup_{j\in M^i}\{P^S_j(t), E^S_j(t), \xi^S_j(t) \}, \\ 
\bigcup_{l\in L^i}\{P^A_l(t), E^A_l(t), \xi^A_l(t)\}\},
\end{multline}
where 
$P_i(t)$, $\xi_i(t)$, $P^S_j(t)$, $E^S_j(t)$, $\xi^S_j(t)$, $P^A_l(t)$, $E^A_l(t)$, and $\xi^A_l(t)$ stand for the position of $G_i$, the capacity of $G_i$, the position of $S_j^i(t)$, the energy state of $S_j^i(t)$, the capacity of $S_j^i(t)$, the position of $A_l^i(t)$, the energy state of $A_l^i(t)$, and the capacity of $A_l^i(t)$.
Here, the positions of $G_i$, $S_j^i$, and $A_l^i$ are specified as 
$P_i(t) = \{p^\phi_i(t), p^\lambda_i(t), p^H_i(t)\}$, 
$P^S_j(t) = \{p^\phi_j(t), p^\lambda_j(t), p^H_j(t), v^S_j(t)\}$,
and 
$P^A_l(t) = \{p^\phi_l(t), p^\lambda_l(t), p^H_l(t), v^A_l(t)\}$,
where $p^\phi_i(t)$, $p^\lambda_i(t)$, and $p^H_i(t)$ denote the latitude, longitude, and altitude of $G_i$. Similarly, $p^\phi_j(t)$, $p^\lambda_j(t)$, $p^H_j(t)$, $v^S_j(t)$, $p^\phi_l(t)$, $p^\lambda_l(t)$, $p^H_l(t)$, and $v^A_l(t)$ represent the latitude of $S_j^i$, the longitude of $S_j^i$, the altitude of $S_j^i$, the velocity vector of $S_j^i$, the latitude of $A_l^i$, the longitude of $A_l^i$, the altitude of $A_l^i$, and the velocity vector of $A_l^i$.

\BfPara{Action}
The action at $t$ is represented as $\mathcal{A}(t) = {[x^i_{j,l}(t)]}$, where $x^i_{j,l}(t) \in \{0, 1\}$.
This indicates whether $G_i$ is available for $S_j^i$ or $A_l^i$ at $t$ or not, and note that the network access service between $G_i$ and NTN device ($S_j^i$ or $A_l^i$) is available when $x^i_j(t)=1$ or $x^i_l(t)=1$ (vice versa).

\BfPara{Reward}
The reward function is outlined in \eqref{eq:obj-2}, with its maximization reliant on the action scheduling $x^i_{j,l}(t)$ made by $G_i$. This reward encompasses both \textit{utility} and \textit{cost} functions.
Fundamentally, the goal is for each GS to orchestrate the scheduling of NTN devices (CubeSats or HALE-UAVs) to enhance the access performance in global SAGIN systems.
Simultaneously, our reward function aims at the reduction of (i) the overall energy usage and (ii) the standard deviation of individual energy levels of CubeSats and HALE-UAVs.
This reward function facilitates the autonomous and cooperative energy management in CubeSat and HALE-UAV.

\subsection{QMARL-based Scheduler Design}

\begin{figure*}
  \centering
  \includegraphics[width=0.9\linewidth]{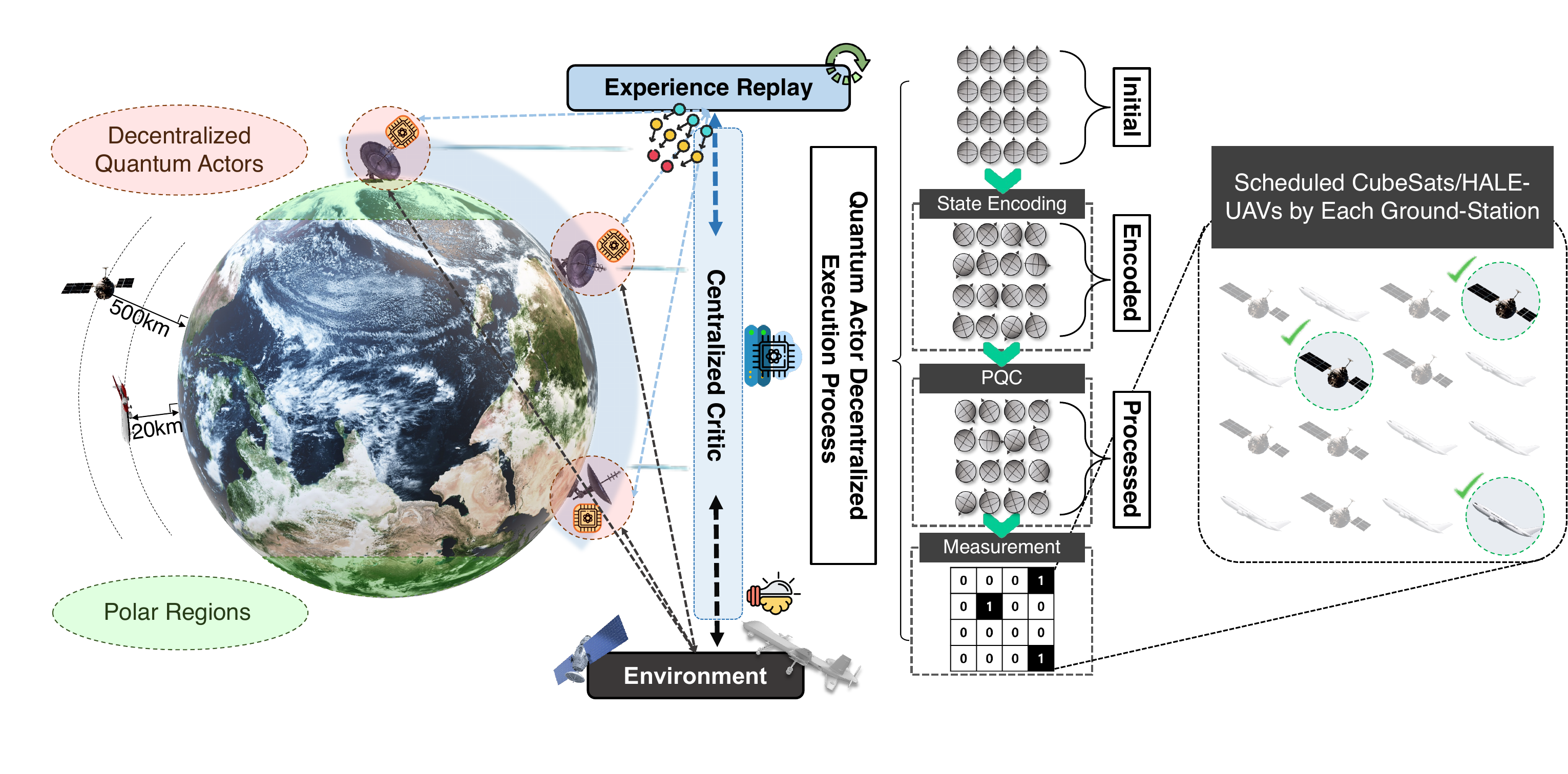}
  \caption{Global SAGIN mobile access using our proposed QMARL-based scheduler.}
  \label{fig:QMARL}
\end{figure*}

In the depicted scenario, each GS agent, identified as the $i$-th GS, is responsible for executing a combinatorial scheduling decision across $M$ CubeSats and $L$ HALE-UAVs, as illustrated in Fig.~\ref{fig:QMARL}.
As the number of CubeSats $M$ and HALE-UAVs $L$ increment linearly, the total number of feasible scheduling decisions experiences an exponential rise, quantified as $2^{M+L}$.
This significant increase highlights the imperative for conventional RL policies to expand their output dimensionality, i.e., \textit{action dimensions}, thereby accommodating the $2^{M+L}$ potential combinations of these scheduling actions.
However, such an increase in output dimensionality introduces difficulties in learning efficacy, a situation often described as the \textit{curse of dimensionality}~\cite{du2021survey}.
To tackle the mentioned challenge, this paper proposes an innovative strategy utilizing QMARL.
This approach leverages quantum measurement techniques, facilitating effective navigation through high-dimensional action decision spaces by GSs.
It's noteworthy that training MARL with a substantial number of agents typically encounters reward convergence issues. 
Furthermore, as the number of action dimensions required by agents rises, achieving reward convergence grows more challenging.
The quantum-based proposed measurement introduced here stands out as a singular solution capable of surmounting these challenges.

The QMARL-based scheduler outlined in this scenario is organized into \textit{three} separate stages.
The first two stages include \textit{encoding}, which involves converting classical bits into quantum states referred to as \textit{qubits}, and \textit{PQC}, which involves the process of applying rotation gates to manipulate these quantum states in accordance with conventional QNN-based RL policies.
The third and most important stage is \textit{measurement}.
During the concluding measurement stage, quantum states are transformed into an \textit{observable}.
This \textit{observable} serves as the output obtained through the measurement of quantum states.
The process of quantum measurement acts as a \textit{decoding} mechanism, translating the outcomes of quantum computing into a format that classical computing systems can interpret and use.
To facilitate global access performance of integrated networks through QMARL, the quantum system is established with a total of $M+L$ qubits. 
This total directly reflects the combined amount of CubeSats ($M$) and HALE-UAVs ($L$), leading to the equation: $|\psi\rangle = \sum_{k=1}^{2^{M+L}} \alpha_{k} |\mathbf{e}_k\rangle$. In this context, $\alpha_{k}$ is defined as the probability amplitude, and $\mathbf{e}_k$ represents the $k$-th basis within the Hilbert space.

In the domain of QNN, the \textit{Pauli-Z} measurement is a prevalent method for transforming quantum states into \textit{observables}.
This conversion process does not depend on the number of qubits in use.
In the \textit{Pauli-Z} operator, each column denotes the computational basis of $|\hat{0}\rangle$ and $|\hat{1}\rangle$.
For the purpose of deriving the expectation value of each qubit's state, a matrix that projects the quantum state onto the $z$-axis is employed, which is expressed as,
${\textbf{\textrm{P}}}^{k}_{Z} \triangleq {\textbf{\textrm{I}}}^{k-1} \otimes {\textbf{\textrm{Z}}} \otimes {\textbf{\textrm{I}}}^{Q-k}$,
where ${\textbf{\textrm{I}}}$ is the identity matrix.
The equation to compute an \textit{observable} associated with a single basis is formulated as, $\langle\mathcal{O}_{k}\rangle = 
\langle\psi| {\textbf{\textrm{P}}}^{k}_{Z} |\psi\rangle$, where $\forall k \in \mathbbm{N}[1,Q]$, $\langle\mathcal{O}_{k}\rangle \in \mathbbm{R}[-1,1]$.
To manage the combinatorial scheduling of $M$ CubeSats and $L$ HALE-UAVs, a requisite output dimensionality of $2^{M+L}$ necessitates the use of $2^{M+L}$ qubits. This methodology, however, does not address the issue identified as the \textit{curse of dimensionality}. 
In contrast, the QMARL-based scheduler proposed in this paper effectively minimizes the requisite number of qubits to a logarithmic scale, transitioning from $2^{M+L}$ down to $M+L$.
Consequently, this innovative approach significantly reduces the qubit requirement, ensuring its operational feasibility even amidst the constraints of the noisy intermediate-scale quantum (NISQ) era, where qubit availability is limited.
By implementing the \textit{basis} measurement, particularly through PVM, the approach outlined in this paper facilitates the determination of probabilities for every possible $2^{M+L}$ combinations with merely $M+L$ qubits. 
Thus, the likelihood of each conceivable $2^{M+L}$ action can be ascertained using only $M+L$ qubits, expressed as,
$\{\text{Pr}(\mathcal{A}_k)\}_{k=1}^{2^{M+L}} \triangleq \{\Motimes_{k=1}^{M+L}\nolimits   |x^i_{j,l}\rangle$\},
where $\Motimes$ symbolizes the \textit{Kronecker product}, $\forall x^i_{j,l} \in \{0,1\}$, $\forall j \in [1,M]$, $\forall l \in [1,L]$.
Finally, the process to determine the probability that the $i$-th GS will choose for the $k$-th action from $2^{M+L}$ possibilities at $t$, according to its strategy, is represented as,
\begin{equation}
\pi(\mathcal{A}_{k}(t)|\mathcal{S}_{i}(t);\boldsymbol{\theta}_{i}) \!=\! \langle\psi|\mathbf{e}_k\rangle\langle \mathbf{e}_k|\psi\rangle \!=\! |\langle\psi|\mathbf{e}_k\rangle|^2 \!=\! |\alpha_{k}|^2,
\end{equation}
where $|\mathbf{e}_k\rangle\langle \mathbf{e}_k|$ denotes the projector for the $k$-th basis, with the collection of all such projectors for every basis being $\{|\mathbf{e}_k\rangle\langle \mathbf{e}_k|\}^{2^{M+L}}_{k=1}$.
This is because the probabilities for each action corresponds to an individual outputs as,
$\sum^{2^{M+L}}_{k=1}\nolimits \pi(\mathcal{A}_{k}(t)|\mathcal{S}_{i}(t); \thetab_i) = 1$.
This paper adopts activation functions as \textit{basis} measurement, thereby allowing each GS to undertake action decision-making on the logarithmically reduced action dimension.

\subsection{QMARL-based Scheduler Training}
The network under consideration is conceptualized as a multi-agent system, where each $i$-th GS acts as the $i$-th agent equipped with its own QNN-based RL policy, $\pi(\mathcal{A}(t)|\mathcal{S}_i(t);\thetab_i)$, parameterized by $\boldsymbol{\theta}_i$.
In the training phase, a unified centralized critic, parameterized by $\phi$, assesses the policy effectiveness of multiple agents by estimating the state-value function $V_{\boldsymbol{\phi}}(\mathcal{S}(t))$, with $\mathcal{S}(t)$ representing the ground truth, encapsulating all accessible environmental data~\cite{lowe2017multi}.
Conversely, each GS engages in sequential decision-making based on its individual partial state (i.e., observation), $\mathcal{S}_i(t)$.
This training framework enables all GSs to refine their policies towards collective decision-making, notwithstanding their limited observation of the environment.
Furthermore, during inference, due to the distributed approach to cooperation, it is possible to achieve effective scalability and efficient use of computing resources.

After completing this procedure, TD error is utilized to implement multi-agent PG methods for the training of quantum multi-actor centralized-critic networks. The objective function for the $i$-th actor ($G_i$), denoted as $J(\thetab_i)$, is expressed as,
\begin{equation}
    \nabla_{\theta_i}J(\theta_i) = \mathbbm{E}_{\mathcal{S}} \Big[\!\sum^{T}_{t=1}\nolimits \sum^{N}_{i=1}\nolimits    
    \delta_{\phi}(t) \nabla_{\theta_i} \!\log\pi(\mathcal{A}(t)|\mathcal{S}_i(t);\theta_i)\Big],
\end{equation}
where $\delta_{\phi}(t)$, $\pi$, $\mathcal{A}(t)$, $\mathcal{S}_i(t)$, and $\theta_i$ are the TD error based on Bellman optimality equation in time step $t$, policy, action at time $t$, state at time $t$, and neural network parameters, respectively.
The loss function pertaining to the critic, denoted by $\mathcal{L}(\phi)$, is specified as,
\begin{equation}
    \nabla_{\phi}\mathcal{L}(\phi) = \sum^{T}_{t=1}\nolimits  \nabla_{\phi}\left\|\delta_{\phi}(t)\right\|^2,
\end{equation}
To optimize the objective function for multiple GSs and reduce the loss function of the centralized critic, the derivatives of the $k$-th parameters are expressed as,
\begin{align}
    & 
    \frac{\partial J(\theta_i)}{\partial \theta_k} = \underbrace{\frac{\partial J(\theta_i)}{\partial \pi_{\theta_i}} \cdot \frac{\partial \pi_{\theta_i}}{\partial \langle \mathcal{O}_{k,\thetab_i} \rangle}}_{\textrm{(Classical Backpropagation)}} \cdot \underbrace{\frac{{\partial \langle \mathcal{O}_{k,\thetab_i} \rangle}}{\partial \theta_k}}_{\textrm{(Parameter-Shift Rule)}}, \label{eq:loss_actor_derivative} \\
    & 
    \frac{\partial\mathcal{L}(\phi)}{\partial \phi_k}  = \underbrace{\frac{\partial\mathcal{L}(\phi)}{\partial V_{\phi}} \cdot \frac{\partial V_{\phi}}{\partial \langle \mathcal{O}_{k,\phi} \rangle}}_{\textrm{(Classical Backpropagation)}}\cdot \underbrace{\frac{{\partial \langle \mathcal{O}_{k,\phi} \rangle}}{\partial \phi_k}}_{\textrm{(Parameter-Shift Rule)}}, \label{eq:loss_critic_derivative}
\end{align}
and the first and second terms of the right-hand side in \eqref{eq:loss_actor_derivative} and \eqref{eq:loss_critic_derivative} are computed using classical partial derivatives.
Nonetheless, the third term presents a challenge for classical computation methods, as the quantum state's specifics remain indeterminate before collapsing its state by measurement.
To overcome this problem in parameter optimization throughout the training phase, the \textit{parameter shift rule} comes into play.
The rule applied for computing the derivative of the $i$-th GS's $k$-th parameter, focusing on the $0$-th order derivative, is specified as,
\begin{equation}
\frac{{\partial \langle \mathcal{O}_{k,\thetab_i} \rangle}}{\partial \theta_k} =\langle \mathcal{O}_{k,\thetab_i + \frac{\pi}{2} \mathbf{e}_k } \rangle - \langle \mathcal{O}_{k,\thetab_i - \frac{\pi}{2} \mathbf{e}_k } \rangle,
\end{equation}
where $\mathbf{e}_k$ denotes the $k$-th basis.
Unlike classical backpropagation, the \textit{parameter shift rule} provides a more straightforward and intuitive methodology.
As a result, this approach can significantly expedite the training process for QNNs.

\section{Performance Evaluation}\label{sec:5_Performance Evaluation}

\subsection{Benchmarks and Simulation Setup}
To evaluate the performance of the \textit{dimension-reduced} QMARL-based scheduler, various benchmarks are utilized, i.e., \textit{MARL}, \textit{Independent Q-Learning (IQL)}, \textit{Deep Q-Network (DQN)}, and \textit{Random} (i.e., Monte Carlo) schedulers.
In the \eqref{eq:quailty_function} for the quality function, $\xi_1$ and $\xi_2$ are $\xi_1=0.01$ and $\xi_2=1,024$, respectively, and
the parameters used for this performance evaluation are presented in Table~\ref{tab:parameter-qmarl}.

\begin{table}[t!]
\centering
\caption{System Parameters for Performance Evaluation}
\scriptsize 
\resizebox{\columnwidth}{!}{ 
\begin{tabular}{l|r}\toprule[1pt]
    \textbf{\textsf{Notation}} & \textbf{\textsf{Value}} \\
    \midrule
    No. of GSs/CubeSats/HALE-UAVs ($N$, $M$, $L$) & $4$, $8$, $8$ \\
    Action dimension ($|\mathcal{A}|$)& $\{2^{1}, 2^{4}, 2^{16}\}$ \\
    Discount factor ($\gamma$)& $0.98$ \\
    Batch size & $64$ \\
    Initial/Min of epsilon ($\epsilon_{\mathrm{init}}$, $\epsilon_{\min}$) & $0.275$, $10^{-2}$ \\
    Annealing epsilon & $5 \times 10^{-5}$ \\
    LR of \textit{actor} ($\alpha_{\mathrm{actor}}$)&  $10^{-3}$ \\
    LR of \textit{central critic} ($\alpha_{\mathrm{critic}}$) & $2.5\times 10^{-4}$ \\
    Training epochs & $10,000$ \\
    Activation & ReLU, Optimizer: Adam \\
    \bottomrule[1pt]
\end{tabular}
}
\label{tab:parameter-qmarl}
\end{table}

\subsection{Policy Training}

\begin{figure*}
    \centering
    \includegraphics[width=0.65\linewidth]{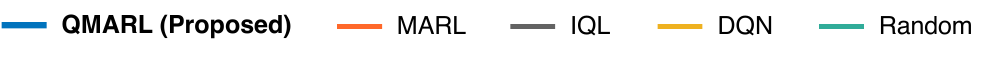}\\
    \subfigure[Reward.]{
    \includegraphics[width=0.30\linewidth]{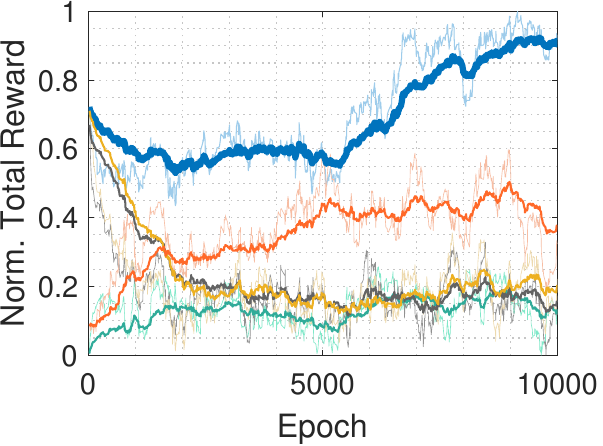}
    \label{fig:reward}
    }
    \subfigure[QoS.]{
    \includegraphics[width=0.30\linewidth]{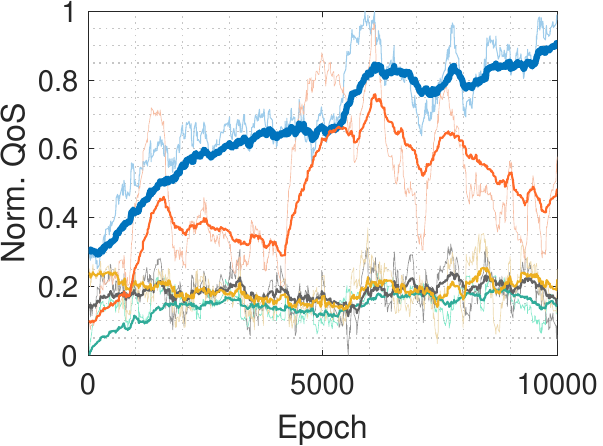}
    \label{fig:QoS}
    }
    \subfigure[Capacity.]{
    \includegraphics[width=0.30\linewidth]{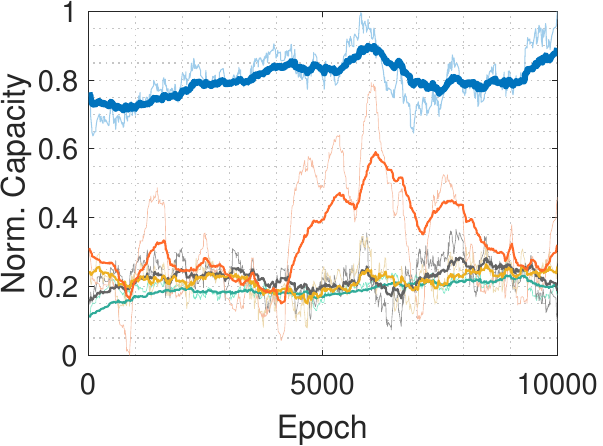}
    \label{fig:capacity}
    }\\
    \subfigure[Residual energy of CubeSat.]{
    \includegraphics[width=0.30\linewidth]{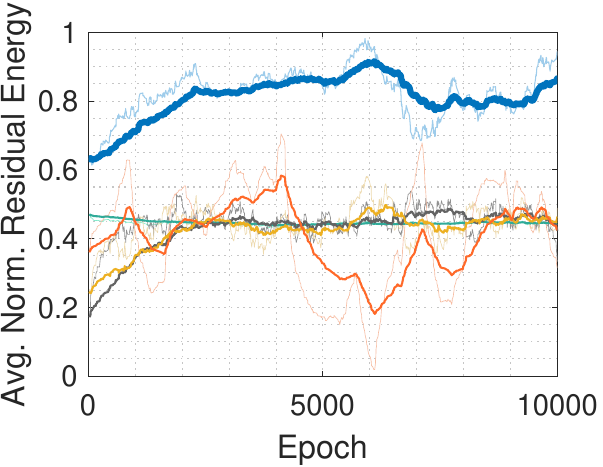}
    \label{fig:energy of CubeSat}
    }
    \subfigure[Residual energy of HALE-UAV.]{
    \includegraphics[width=0.30\linewidth]{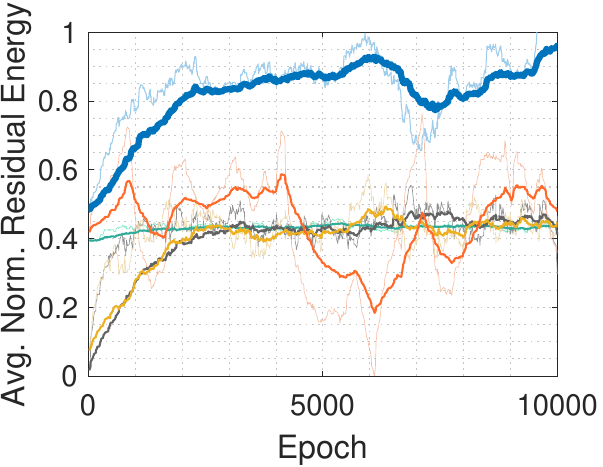}
    \label{fig:energy of HALE-UAV}
    } 
    \caption{SAGIN access performance, i.e., access availability (QoS, capacity) and energy efficiency (residual energy).}
    \label{fig:training_result}
\end{figure*}

Fig.~\ref{fig:reward} illustrate that the QMARL-based scheduling approach introduced in this paper outperforms comparative benchmarks, achieving a maximal reward of $1.0$. In comparison, the MARL-based scheduler provides less reward than the QMARL-based scheduler, and the reward value fluctuates and eventually does not converge.
Furthermore, the performance of IQL and DQN based schedulers closely mirrors that of the Random based scheduler in terms of reward. Figs.~\ref{fig:training_result}(b)-(e) reveal that the scheduler based on QMARL attains superior QoS, capacity, and remaining energy for CubeSats/HALE-UAVs. Conversely, MARL-based scheduling approaches fail to concurrently optimize multiple metrics related to communication and the energy efficiency of NTN devices. Within the MARL based-scheduler, an increase in QoS and capacity correlates with a decrease in residual energy, indicating an inability to simultaneously optimize global access performance of integrated networks (QoS, capacity) and the residual energy of CubeSats/HALE-UAVs. In contrast, the QMARL-based scheduler successfully optimizes both global access performance and energy efficiency in parallel.
\begin{table}[t!]
\centering
\scriptsize 
\caption{Performance Evaluation Results when $|\mathcal{A}| = 2^{16}$}
\resizebox{\columnwidth}{!}{ 
\begin{tabularx}{1\linewidth}{l c c c}
    \toprule[1pt]

Algorithm & QoS  & Capacity & Residual Energy \\
\cmidrule(lr){1-1} \cmidrule(lr){2-2} \cmidrule(lr){3-3} \cmidrule(lr){4-4} 

\textbf{QMARL}
& $\mathbf{0.906}$\; \tikz{
\draw[gray,line width=.3pt] (0,0) -- (1.1,0);
\draw[white, line width=0.01pt] (0,-2pt) -- (0,2pt);
\draw[black,line width=1pt] (0.906,0) -- (0.974,0);
\draw[black,line width=1pt] (0.906,-2pt) -- (0.906,2pt);
\draw[black,line width=1pt] (1.000,-2pt) -- (1.000,2pt);}
& $\mathbf{0.894}$\; \tikz{
\draw[gray,line width=.3pt] (0,0) -- (1.1,0);
\draw[white, line width=0.01pt] (0,-2pt) -- (0,2pt);
\draw[black,line width=1pt] (0.894,0) -- (0.994,0);
\draw[black,line width=1pt] (0.894,-2pt) -- (0.894,2pt);
\draw[black,line width=1pt] (0.994,-2pt) -- (0.994,2pt);}
& $\mathbf{0.912}$\; \tikz{
\draw[gray,line width=.3pt] (0,0) -- (1.1,0);
\draw[white, line width=0.01pt] (0,-2pt) -- (0,2pt);
\draw[black,line width=1pt] (0.912,0) -- (1.000,0);
\draw[black,line width=1pt] (0.912,-2pt) -- (0.912,2pt);
\draw[black,line width=1pt] (1.000,-2pt) -- (1.000,2pt);} \\

MARL
& $0.484$\; \tikz{
\draw[gray,line width=.3pt] (0,0) -- (1.1,0);
\draw[white, line width=0.01pt] (0,-2pt) -- (0,2pt);
\draw[black,line width=1pt] (0.484,0) -- (0.584,0);
\draw[black,line width=1pt] (0.484,-2pt) -- (0.484,2pt);
\draw[black,line width=1pt] (0.584,-2pt) -- (0.584,2pt);}
& $0.321$\; \tikz{
\draw[gray,line width=.3pt] (0,0) -- (1.1,0);
\draw[white, line width=0.01pt] (0,-2pt) -- (0,2pt);
\draw[black,line width=1pt] (0.321,0) -- (0.421,0);
\draw[black,line width=1pt] (0.321,-2pt) -- (0.321,2pt);
\draw[black,line width=1pt] (0.421,-2pt) -- (0.421,2pt);}
& $0.457$\; \tikz{
\draw[gray,line width=.3pt] (0,0) -- (1.1,0);
\draw[white, line width=0.01pt] (0,-2pt) -- (0,2pt);
\draw[black,line width=1pt] (0.457,0) -- (0.557,0);
\draw[black,line width=1pt] (0.457,-2pt) -- (0.457,2pt);
\draw[black,line width=1pt] (0.557,-2pt) -- (0.557,2pt);} \\

IQL
& $0.148$\; \tikz{
\draw[gray,line width=.3pt] (0,0) -- (1.1,0);
\draw[white, line width=0.01pt] (0,-2pt) -- (0,2pt);
\draw[black,line width=1pt] (0.148,0) -- (0.248,0);
\draw[black,line width=1pt] (0.148,-2pt) -- (0.148,2pt);
\draw[black,line width=1pt] (0.248,-2pt) -- (0.248,2pt);}
& $0.188$\; \tikz{
\draw[gray,line width=.3pt] (0,0) -- (1.1,0);
\draw[white, line width=0.01pt] (0,-2pt) -- (0,2pt);
\draw[black,line width=1pt] (0.188,0) -- (0.288,0);
\draw[black,line width=1pt] (0.188,-2pt) -- (0.188,2pt);
\draw[black,line width=1pt] (0.288,-2pt) -- (0.288,2pt);}
& $0.419$\; \tikz{
\draw[gray,line width=.3pt] (0,0) -- (1.1,0);
\draw[white, line width=0.01pt] (0,-2pt) -- (0,2pt);
\draw[black,line width=1pt] (0.419,0) -- (0.519,0);
\draw[black,line width=1pt] (0.419,-2pt) -- (0.419,2pt);
\draw[black,line width=1pt] (0.519,-2pt) -- (0.519,2pt);} \\

DQN
& $0.194$\; \tikz{
\draw[gray,line width=.3pt] (0,0) -- (1.1,0);
\draw[white, line width=0.01pt] (0,-2pt) -- (0,2pt);
\draw[black,line width=1pt] (0.194,0) -- (0.294,0);
\draw[black,line width=1pt] (0.194,-2pt) -- (0.194,2pt);
\draw[black,line width=1pt] (0.294,-2pt) -- (0.294,2pt);}
& $0.258$\; \tikz{
\draw[gray,line width=.3pt] (0,0) -- (1.1,0);
\draw[white, line width=0.01pt] (0,-2pt) -- (0,2pt);
\draw[black,line width=1pt] (0.258,0) -- (0.358,0);
\draw[black,line width=1pt] (0.258,-2pt) -- (0.258,2pt);
\draw[black,line width=1pt] (0.358,-2pt) -- (0.358,2pt);}
& $0.442$\; \tikz{
\draw[gray,line width=.3pt] (0,0) -- (1.1,0);
\draw[white, line width=0.01pt] (0,-2pt) -- (0,2pt);
\draw[black,line width=1pt] (0.442,0) -- (0.542,0);
\draw[black,line width=1pt] (0.442,-2pt) -- (0.442,2pt);
\draw[black,line width=1pt] (0.542,-2pt) -- (0.542,2pt);} \\

Random
& $0.151$\; \tikz{
\draw[gray,line width=.3pt] (0,0) -- (1.1,0);
\draw[white, line width=0.01pt] (0,-2pt) -- (0,2pt);
\draw[black,line width=1pt] (0.151,0) -- (0.251,0);
\draw[black,line width=1pt] (0.151,-2pt) -- (0.151,2pt);
\draw[black,line width=1pt] (0.251,-2pt) -- (0.251,2pt);}
& $0.197$\; \tikz{
\draw[gray,line width=.3pt] (0,0) -- (1.1,0);
\draw[white, line width=0.01pt] (0,-2pt) -- (0,2pt);
\draw[black,line width=1pt] (0.197,0) -- (0.297,0);
\draw[black,line width=1pt] (0.197,-2pt) -- (0.197,2pt);
\draw[black,line width=1pt] (0.297,-2pt) -- (0.297,2pt);}
& $0.437$\; \tikz{
\draw[gray,line width=.3pt] (0,0) -- (1.1,0);
\draw[white, line width=0.01pt] (0,-2pt) -- (0,2pt);
\draw[black,line width=1pt] (0.437,0) -- (0.537,0);
\draw[black,line width=1pt] (0.437,-2pt) -- (0.437,2pt);
\draw[black,line width=1pt] (0.537,-2pt) -- (0.537,2pt);} \\

\bottomrule[1pt]
\end{tabularx}
}
\label{tab:training_result}
\end{table}

Table~\ref{tab:training_result} illustrates that the QMARL based scheduler significantly surpasses its MARL-based scheduler, recording an 87.2$\%$ enhancement in QoS, a 178$\%$ increase in capacity, and an 99.5$\%$ augmentation in remaining energy.
Additionally, the performance of IQL, DQN, and Random based scheduler are notably inferior in all evaluated aspects, with QoS not exceeding $0.2$, capacity remaining below $0.26$, and the residual energy of CubeSats/HALE-UAVs falling short of $0.45$, as explicated in Table~\ref{tab:training_result}.

\begin{figure*}[t!]
    \centering
    \includegraphics[width=0.65\linewidth]{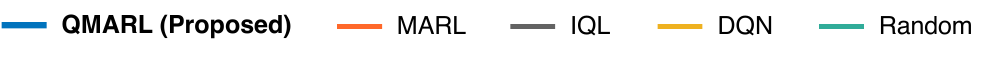}\\
    \subfigure[QoS vs. Residual Energy.]{
    \includegraphics[width=0.33\linewidth]{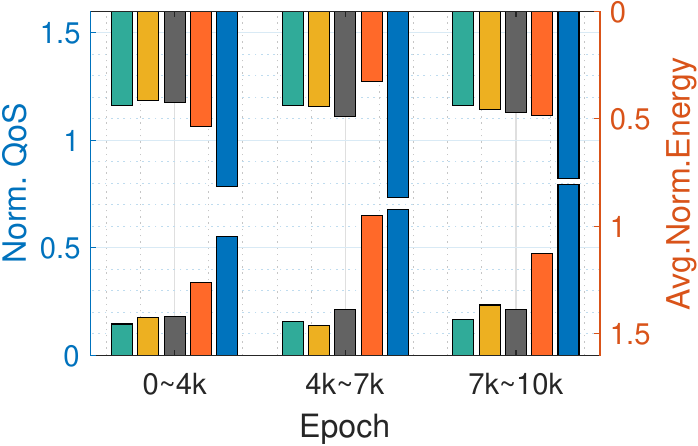}
    }
    \subfigure[Capacity vs. Residual Energy.]{
    \includegraphics[width=0.33\linewidth]{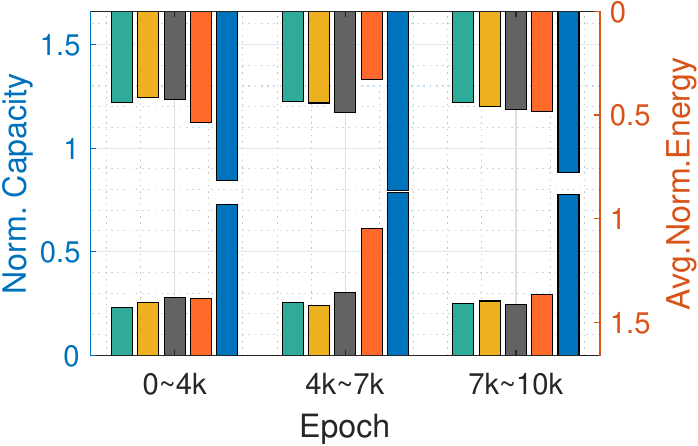}
    } \\
    \subfigure[Residual energy for each CubeSat.]{
    \includegraphics[width=0.33\linewidth]{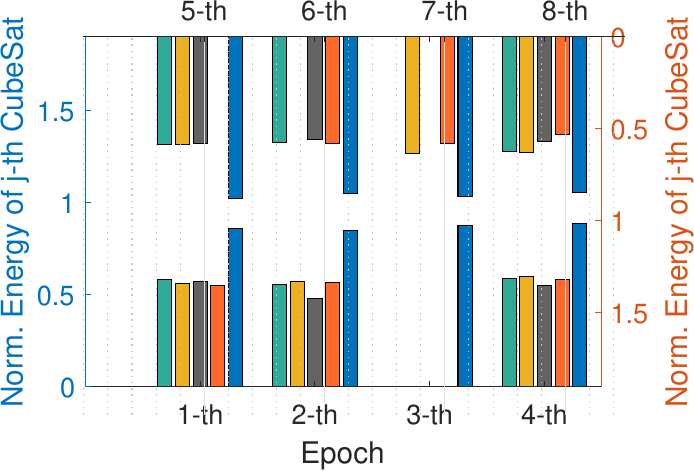}
    } 
    \subfigure[Residual energy for each HALE-UAV.]{
    \includegraphics[width=0.33\linewidth]{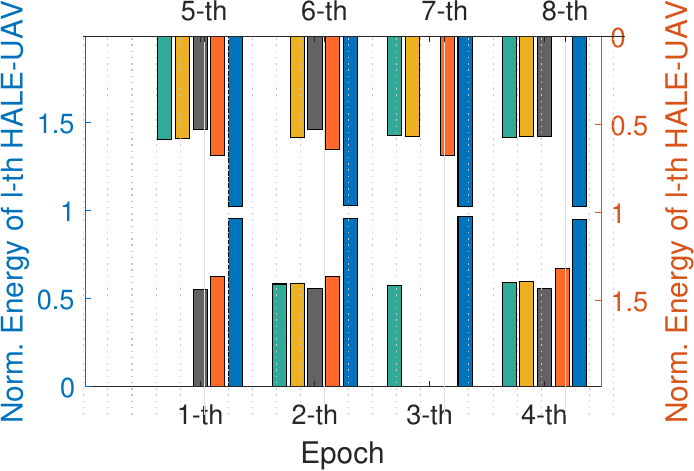}
    }
    \caption{Relationship between access availability and energy efficiency.}
    \label{fig:training_result_bar}
\end{figure*}

Figs.~\ref{fig:training_result_bar}(a)--(b) delineate the correlation between the global access performance of integrated networks and the normalized residual energy of NTNs, contingent upon the employed algorithm.
The epoch on the $x$-axis is segmented into three phases: $0$ to $4k$ (initial phase), $4k$ to $7k$ (intermediate phase), and $7k$ to $10k$ (final phase).
Throughout the progression from the initial to the intermediate phase in MARL, an increment is observed in the energy of NTN devices, albeit with a reduction in QoS and capacity.
This limitation is not exclusive to MARL but also extends to schedulers based on IQL, DQN, and Random schedulers, which are unable to concurrently optimize the performance of global access performance of integrated networks and the residual energy of NTN devices.
In stark contrast, QMARL-based scheduler consistently maintains elevated levels of QoS, capacity, and residual energy.
Figs.~\ref{fig:training_result_bar}(c)--(d) display the remaining energy of the $S^i_j$ and $A^i_l$.
The occurrence of non-operational NTN devices is attributed to the inefficiency in energy utilization by the benchmarks, including those based on MARL, IQL, DQN, and Random based schedulers.
In contrast, the QMARL based scheduler consistently exhibits superior residual energy performance, ensuring the avoidance of any non-functional NTN devices.
Additionally, the QMARL-based scheduler has higher residual energy of NTN devices compared to other benchmarks.
\begin{table}[t!]
    \small
    \centering
    \caption{Total Normalized Converged Rewards}
    \begin{tabular}{l||c|c|c|c|c}
    \toprule
    $|\mathcal{A}|$ & \textbf{QMARL} & {MARL} & {IQL} & {DQN} & {Random}\\
    \midrule
    $2^{1}$ & 0.9971 & \textbf{1.0000} & 0.9411 & 0.9527 & 0.2755 \\
    $2^{4}$  & 0.9813 & \textbf{1.0000} & 0.8267 & 0.9215 & 0.5452 \\
    $\mathbf{2^{16}}$  & \textbf{1.0000} & 0.4103 & 0.1730 & 0.2235 & 0.1390 \\
    \bottomrule
    \end{tabular}
    \label{tab:average_reward}
\end{table}

\begin{figure*}[t!]
    \centering
    \includegraphics[width=0.65\linewidth]{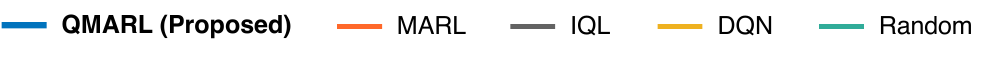}\\
    \subfigure[Distribution of reward values according to action dimensions $|\mathcal{A}|$.]{
    \includegraphics[width=0.33\linewidth]{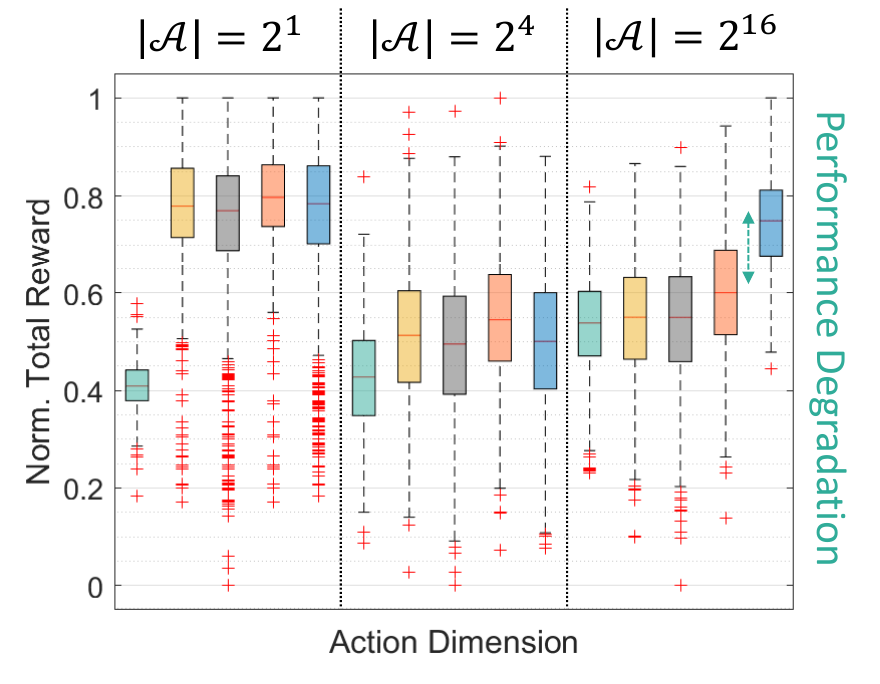}
    \label{fig:box plot}
    }
    \subfigure[Converged rewards according to action dimensions $|\mathcal{A}|$.]{
    \includegraphics[width=0.3\linewidth]{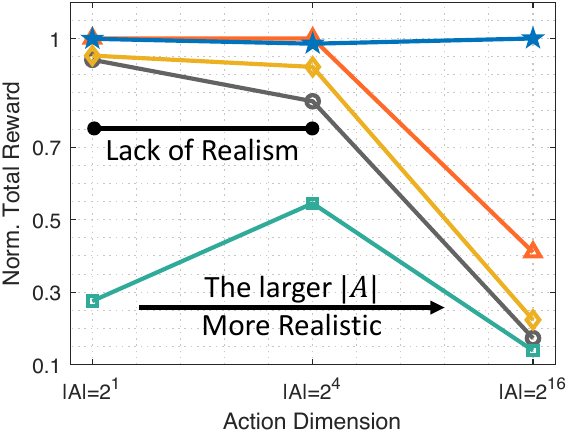}
    \label{fig:converged reward}
    }
    \subfigure[Normalized average residual energy of NTN devices with and without GS-specific capacity requirements.]{
    \includegraphics[width=0.3\linewidth]{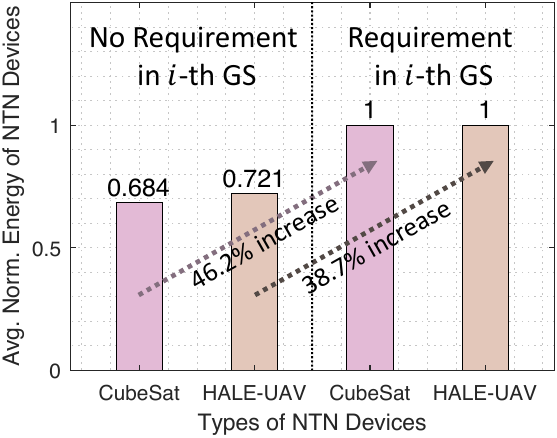}
    \label{fig:bar plot}
    }
    \caption{Rewards due to action dimension and residual energy with or without the capacity requirements in each GS.}
    \label{fig:action_dimension}
\end{figure*}

Figs.~\ref{fig:action_dimension}(a)-(b) and Table \ref{tab:average_reward} provide a comparative analysis of the rewards obtained by GSs utilizing both the proposed algorithms and benchmarks across varying sizes of the action dimension, specifically for $|\mathcal{A}| \in \{2^1$, $2^4$, $2^{16}\}$.
The MARL-based scheduler exhibits superior reward outcomes at smaller action dimensions ($|\mathcal{A}| \in \{2^1, 2^4\}$); however, it encounters significant difficulties at larger action dimension ($|\mathcal{A}| = 2^{16}$), where its performance falls behind that of the QMARL based scheduler by 41.03$\%$, due to the \textit{curse of dimensionality}.
In a similar vein, IQL, DQN based schedulers yield outcomes that are analogous to those of a Random based scheduler at the largest action dimension ($|\mathcal{A}| = 2^{16}$).
Fig.~\ref{fig:box plot} depicts a box plot summarizing the reward distribution across all action dimensions throughout the training process. 
The median reward is represented by the red line at the center of each box, with the lower and upper boundaries of the box indicating the 25$\%$ and 75$\%$, respectively. Outliers are marked with a red '+' symbol.
Notably, at the exceedingly large action dimension ($|\mathcal{A}| = 2^{16}$), the QMARL-based scheduler achieves the highest reward, while the performance of other benchmarks deteriorates.
Fig.~\ref{fig:converged reward} illustrates the converged normalized reward values according to the action dimensions.
The utilization of larger action dimensions is deemed more realistic due to the inclusion of a greater number of CubeSats and HALE-UAVs, hence enhancing real-world applicability.
In global access of integrated networks involving extensive deployment of CubeSats and HALE-UAVs, solely the QMARL-based scheduler achieves successful training outcomes, thereby evidencing a significant performance disparity in comparison to other benchmarks.
These training results distinctly emphasize the exceptional capability of the QMARL based scheduler in addressing and mitigating the challenges posed by the \textit{curse of dimensionality}.

Additionally, Fig.~\ref{fig:bar plot} shows the normalized average residual energy of NTN devices with and without GS-specific capacity requirements.
The pink bar graph represents the average residual energy of CubeSats, and the beige bar graph represents the average residual energy of HALE-UAVs.
In addition, the two bar graphs on the left are when there are no capacity requirements for each GS, and the two bar graphs on the right are when there are capacity requirements for each GS.
If there are capacity requirements for each GS, unnecessary energy waste in NTN devices can be prevented. If the maximum capacity requirements are set differently for each GS depending on the region where the GS is located, the population of the region, and the degree of communication overload, the residual energy for CubeSat is 46.2$\%$ and HALE-UAV is 38.7$\%$ higher.

\section{Concluding Remarks}\label{sec:6_Concluding Remarks}

This paper introduces a novel QMARL-based global SAGIN mobile access scheduler for CubeSats and HALE-UAVs, which aims at the maximization of access availability and energy efficiency. The CubeSats, characterized by their limited energy resources, employ energy efficiency strategies that differentiate between sun side and dark side orbital segments to conserve power. The reason why the quantum-based approach is utilized is that it can realize scheduling action dimension reduction.
This attribute is particularly advantageous for ensuring the robust convergence of rewards in scenarios entailing extensive-scale actions, such as global access with considerable numbers of CubeSats and HALE-UAVs. The study's experimental setup reflects real-world conditions by incorporating the orbital dynamics of CubeSats and the aerodynamic characteristics of HALE-UAVs, thereby underscoring the practical applicability of our proposed QMARL-based scheduler. 
Our performance evaluations with various aspects and benchmarks verify that our proposed scheduler can achieve desired performance improvements. 

{
    \bibliographystyle{IEEEtran}  

}

\newpage

\begin{IEEEbiography}[{\includegraphics[width=1in,height=1.25in,clip,keepaspectratio]{author1.jpg}}]{Gyu Seon Kim} 
is currently a Ph.D. student at the Department of Electrical and Computer Engineering, Korea University, Seoul, Republic of Korea. He received a B.S. degree in aerospace engineering from Inha University, Incheon, Republic of Korea. His research focuses include deep reinforcement learning algorithms and their applications to autonomous mobility systems.

He received the IEEE Seoul Section Student Paper Contest Award (2023).
\end{IEEEbiography}

\begin{IEEEbiography}[{\includegraphics[width=1in,height=1.25in,clip,keepaspectratio]{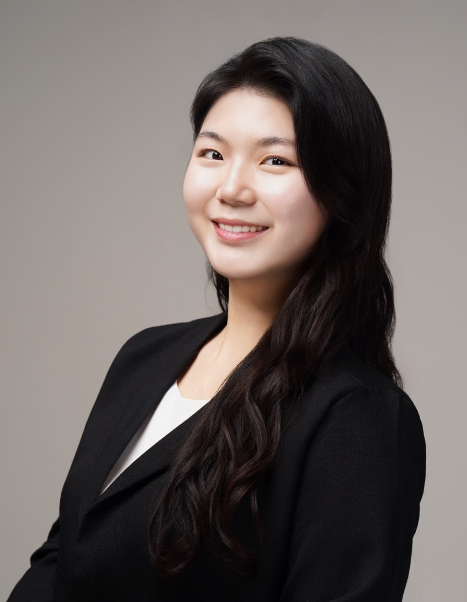}}]{Yeryeong Cho} 
is currently an M.S. student at the Department of Electrical and Computer Engineering, Korea University, Seoul, Republic of Korea. She received a B.S. degree in Robotics \& Convergence from Hanyang University, Ansan, Republic of Korea. She was with the Eco-friendly Smart System Technical Research Center, Incheon,
Republic of Korea, from 2020 to 2022. Her research focuses include deep reinforcement learning algorithms and their applications to autonomous mobility systems.
\end{IEEEbiography}

\begin{IEEEbiography}[{\includegraphics[width=1in,height=1.25in,clip,keepaspectratio]{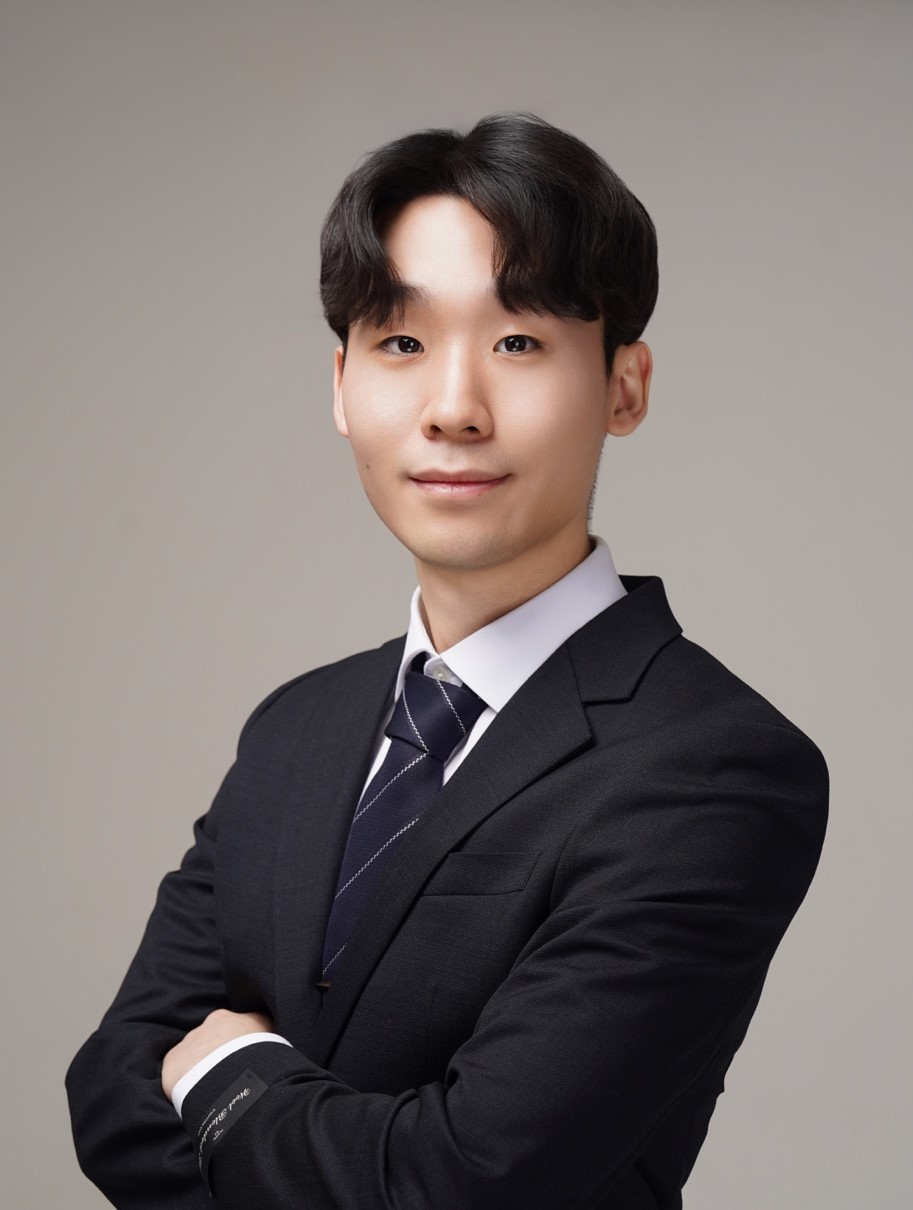}}]{Jaehyun Chung} 
is currently an M.S. student at the Department of Electrical and Computer Engineering, Korea University, Seoul, Korea, where he received his B.S. in electrical engineering, in August 2023.
\end{IEEEbiography}

\begin{IEEEbiography}[{\includegraphics[width=1in,height=1.25in,clip,keepaspectratio]{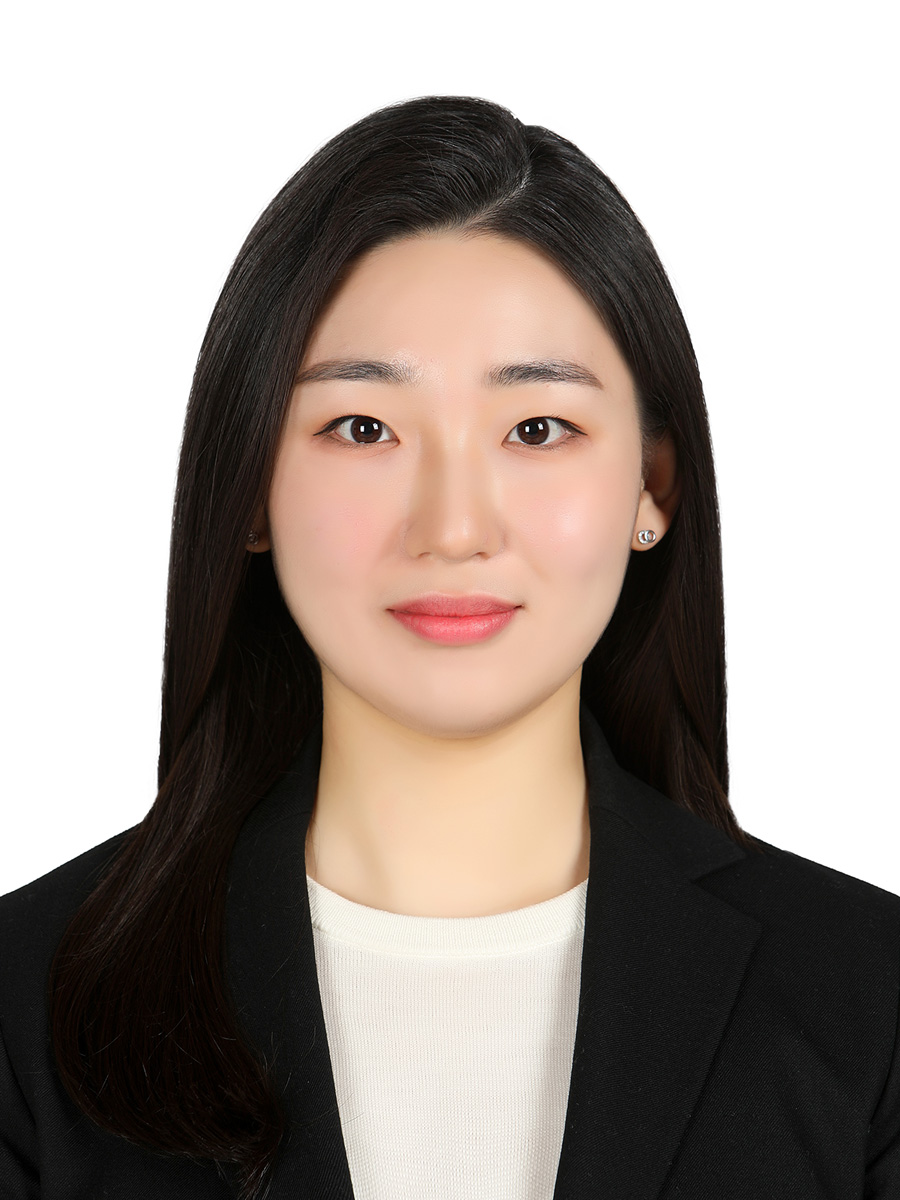}}]{Soohyun Park} 
(Member, IEEE) 
has been an assistant professor at Sookmyung Women's University, Seoul, Korea, since March 2024. She was a postdoctoral scholar at the Department of Electrical and Computer Engineering, Korea University, Seoul, Korea, from September 2023 to February 2024, where she received her Ph.D. in electrical and computer engineering, in August 2023. She also received her B.S. in computer science and engineering from Chung-Ang University, Seoul, Korea, in February 2019. 
She was a recipient of ICT Express Best Reviewer Award (2021), IEEE Seoul Section Student Paper Contest Awards, and IEEE Vehicular Technology Society (VTS) Seoul Chapter Awards.
\end{IEEEbiography}

\begin{IEEEbiography}[{\includegraphics[width=1in,height=1.25in,clip,keepaspectratio]{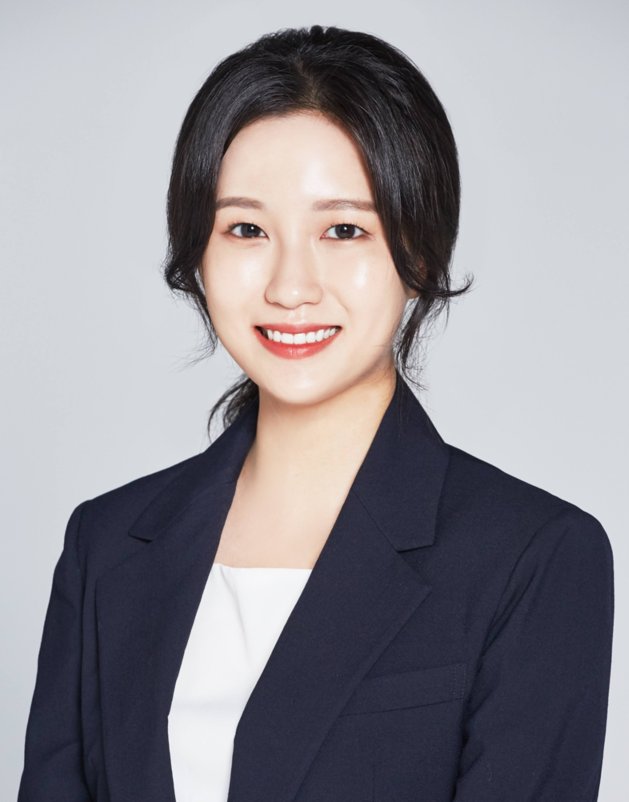}}]{Soyi Jung} 
(Member, IEEE) has been an assistant professor at Ajou University, Suwon, Korea, since September 2022. Before joining Ajou University, she was an assistant professor at Hallym University, Chuncheon, Korea, from 2021 to 2022; a visiting scholar at Donald Bren School of Information and Computer Sciences, University of California, Irvine, CA, USA, from 2021 to 2022; a research professor at Korea University, Seoul, Korea, in 2021; and a researcher at Korea Testing and Research (KTR) Institute, Gwacheon, Korea, from 2015 to 2016. She received her B.S., M.S., and Ph.D. degrees in electrical and computer engineering from Ajou University, Suwon, Korea, in 2013, 2015, and 2021. 
She was a recipient of IEEE Seoul Section Student Paper Contest Award (2018) and IEEE ICOIN Best Paper Award (2021).
\end{IEEEbiography}

\begin{IEEEbiography}[{\includegraphics[width=1in,height=1.25in,clip,keepaspectratio]{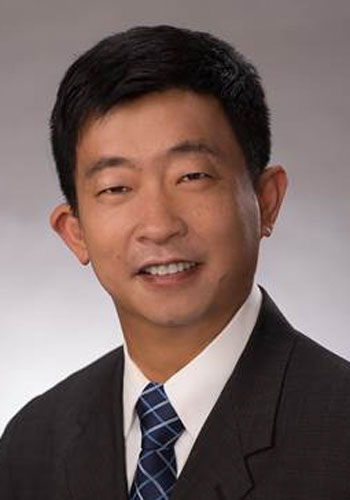}}]{Zhu Han} (Fellow, IEEE) received the B.S. degree in electronic engineering from Tsinghua University, Beijing, China, in 1997, and the M.S. and Ph.D. degrees in electrical and computer engineering from the University of Maryland at College Park, College Park, MD, USA, in 1999 and 2003, respectively. From 2000 to 2002, he was a Research and Development Engineer with JDSU, Germantown, MD, USA. From 2003 to 2006, he was a Research Associate with the University of Maryland at College Park. From 2006 to 2008, he was an Assistant Professor with Boise State University, Boise, ID, USA. He is currently a John and
Rebecca Moores Professor with the Electrical and Computer Engineering Department as well as the Computer Science Department, University of Houston, Houston, TX, USA. He also works with the Department of Computer Science and Engineering, Kyung Hee University, Seoul, South Korea. His
main research targets on the novel game-theory-related concepts critical to enabling efficient and distributive use of wireless networks with limited resources. His other research interests include wireless resource allocation and management, wireless communications and networking, quantum
computing, data science, smart grid, carbon neutralization, security, and privacy. Dr. Han received the NSF Career Award in 2010, the Fred W. Ellersick Prize of the IEEE Communication Society in 2011, the EURASIP Best Paper Award
for the Journal on Advances in Signal Processing in 2015, the IEEE Leonard G. Abraham Prize in the field of Communications Systems (Best Paper Award in IEEE JSAC) in 2016, and several best paper awards in IEEE conferences. He was an IEEE Communications Society Distinguished Lecturer from 2015
to 2018 and has been an AAAS Fellow since 2019 and an ACM Distinguished Member since 2019. He has been a 1\% Highly Cited Researcher since 2017 according to Web of Science. He is also the winner of the 2021 IEEE Kiyo Tomiyasu Award (an IEEE Field Award), for outstanding early to mid-career contributions to technologies holding the promise of innovative applications, with the following citation: “for contributions to game theory and distributed management of autonomous communication networks.”
\end{IEEEbiography}

\begin{IEEEbiography}[{\includegraphics[width=1in,height=1.25in,clip]{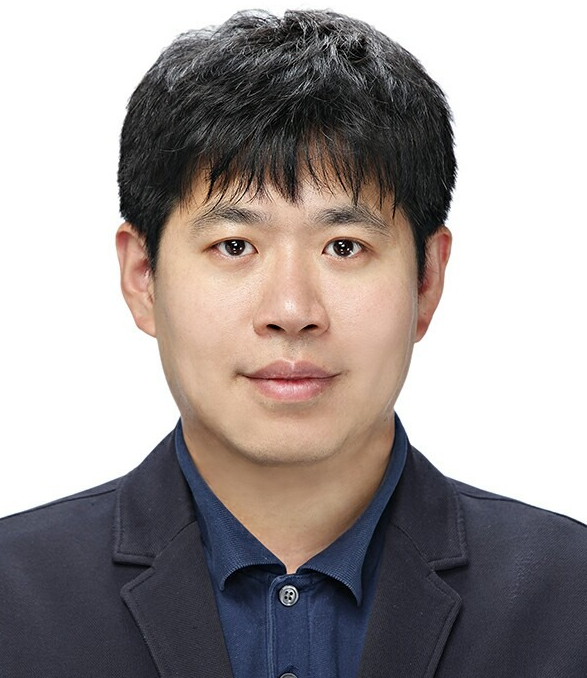}}]{Joongheon Kim} 
(M'06--SM'18) has been with Korea University, Seoul, Korea, since 2019, where he is currently an associate professor at the School of Electrical Engineering. He received the B.S. and M.S. degrees in computer science and engineering from Korea University, Seoul, Korea, in 2004 and 2006; and the Ph.D. degree in computer science from the University of Southern California (USC), Los Angeles, CA, USA, in 2014. Before joining Korea University, he was a research engineer with LG Electronics (Seoul, Korea, 2006--2009), a systems engineer with Intel Corporation (Santa Clara, CA, USA, 2013--2016), and an assistant professor with Chung-Ang University (Seoul, Korea, 2016--2019). 

He serves as an editor for \textsc{IEEE Transactions on Vehicular Technology} and \textsc{IEEE Internet of Things Journal}. 
He was a recipient of Annenberg Graduate Fellowship from USC (2009), 
Intel Corporation Next Generation and Standards (NGS) Division Recognition Award (2015), 
\textsc{IEEE Systems Journal} Best Paper Award (2020), IEEE ComSoc Multimedia Communications Technical Committee (MMTC) Outstanding Young Researcher Award (2020), and IEEE ComSoc MMTC Best Journal Paper Award (2021). He also received IEEE ICOIN Best Paper Award (2021), IEEE ICTC Best Paper Award (2022), and IEEE Vehicular Technology Society (VTS) Seoul Chapter Awards. 
\end{IEEEbiography}

\end{document}